\documentclass{iopjournal}

\usepackage{hyperref}
\usepackage{subcaption}
\usepackage{makecell}
\usepackage{amsmath}

\renewcommand{\articletype}[1]{}

\begin{document}

\articletype{Paper} %	 e.g. Paper, Letter, Topical Review...

\title{Characterization of integrating-sphere systematic effects for photon calibration of the KAGRA gravitational wave detector}

\author{
Shingo Hido$^{1,2,*}$, % orcid{0009-0009-0004-4170}
Misato Onishi$^3$,
Shinji Miyoki$^2$, % orcid{0000-0002-1213-8416}
Takahiro Sawada$^2$, % orcid{0000-0001-5726-7150}
Takahiro Yamamoto$^2$, % orcid{0000-0002-0808-4822}
Takayuki Tomaru$^4$, % orcid{0000-0002-8927-9014}
Yoshiki Moriwaki$^3$, % orcid{0000-0002-4497-6908}
and Dan Chen$^{4,*}$ % orcid{0000-0003-1433-0716}
}

\affil{$^1$Department of Physics, The University of Tokyo, 7-3-1 Hongo, Bunkyo-ku, Tokyo 113-0033, Japan}

\affil{$^2$KAGRA Observatory, Institute for Cosmic Ray Research, The University of Tokyo, 238 Higashi-Mozumi, Kamioka-cho, Hida City, Gifu 506-1205, Japan}

\affil{$^3$Faculty of Science, University of Toyama, 3190 Gofuku, Toyama City, Toyama 930-8555, Japan}

\affil{$^4$Gravitational Wave Science Project, National Astronomical Observatory of Japan, 2-21-1 Osawa, Mitaka, Tokyo 181-8588, Japan}

\affil{$^*$Authors to whom any correspondence should be addressed.}

\email{shingo@icrr.u-tokyo.ac.jp, dan.chen@nao.ac.jp}


\begin{abstract}%
\par Photon calibration (Pcal) systems are essential for accurately reconstructing gravitational-wave strain in laser interferometric detectors.
In the KAGRA Pcal system, the laser power applied to the end test mass is measured using integrating-sphere power sensors.
During the first part of the fourth observing run (O4a), repeated calibration measurements showed scatter larger than expected from the uncertainties assigned to individual measurements.
Although this excess scatter was empirically reflected in the previously reported calibration uncertainties, its physical origin was not identified.
In this study, we experimentally characterize the response of the integrating spheres used in the KAGRA calibration chain with respect to three incident-beam conditions: incidence angle, polarization, and beam position.
Based on these measurements and the reproducibility of the integrating-sphere placement in each calibration configuration, we evaluate the corresponding systematic uncertainties and explicitly incorporate them into the uncertainty model for the O4a Pcal-X calibration measurements.
The inclusion of these previously unmodeled effects substantially improves the consistency between the estimated uncertainties and the observed scatter, particularly for calibration parameters measured at the KAGRA end station, where the reduced chi-square values decrease from values as large as $10^3$ to values of order unity to several.
These results indicate that variations in the incident-beam conditions of integrating-sphere power sensors account for a substantial part of the previously unexplained scatter and establish a more physically motivated framework for evaluating Pcal laser-power calibration uncertainty. They also provide a basis for future reduction of the calibration uncertainty through improved control of the incident-beam conditions.
\end{abstract}

%\subjectindex{xxxx, xxx}

%\maketitle

\section{Introduction}
\label{sec:intro}
    \par Gravitational-wave (GW) observations have opened a new window onto the Universe since the first direct detection by the Laser Interferometer Gravitational-Wave Observatory (LIGO) in 2015~\cite{PhysRevLett.116.061102}.
    The global network of GW detectors, consisting of LIGO in the United States, Virgo in Italy, and KAGRA in Japan, has significantly increased the number of detected events and improved the accuracy of source parameter estimation~\cite{Abac_2026,GWTC5,PhysRevLett.119.141101}.
    The fourth observing run (O4) of the LIGO--Virgo--KAGRA network began in May 2023, and KAGRA participated in the first observing period, O4a, for approximately four weeks~\cite{Chen_2025}.
    As the sensitivity of these detectors continues to improve, the accuracy and reliability of calibration become increasingly important because calibration directly affects the reconstruction of the GW strain data. %cite calがより重要になる系
    
    \par Calibration of interferometric GW detectors is performed by injecting reference signals into the detector, measuring the detector response, and comparing it with models of the sensing and actuation systems to determine and update the model parameters used for reconstructing the GW strain~\cite{Viets_2018,10.1093/ptep/ptab018,Chen:2023fb}.
    In KAGRA, the interferometric configuration during O4a was based on a power-recycled Fabry--Pérot Michelson interferometer (PRFPMI)~\cite{Chen_2025}.
    The interferometer consists of multiple suspended mirrors forming optical cavities, including input test masses (ITMs), end test masses (ETMs), and recycling mirrors.

    \par In current large-scale interferometric GW detectors, photon calibration (Pcal) systems are used as primary reference signal generators for calibration~\cite{10.1063/1.4967303,Estevez_2021,Bhattacharjee_2021}.
    A Pcal system induces calibrated displacements of an ETM through the radiation pressure of auxiliary laser beams, enabling absolute calibration of the detector response.
    Therefore, the accuracy of the Pcal system directly contributes to, and can ultimately limit, the accuracy of the reconstructed GW strain.
    \par
\begin{figure}
        \centering
        \includegraphics[width=0.75\linewidth]{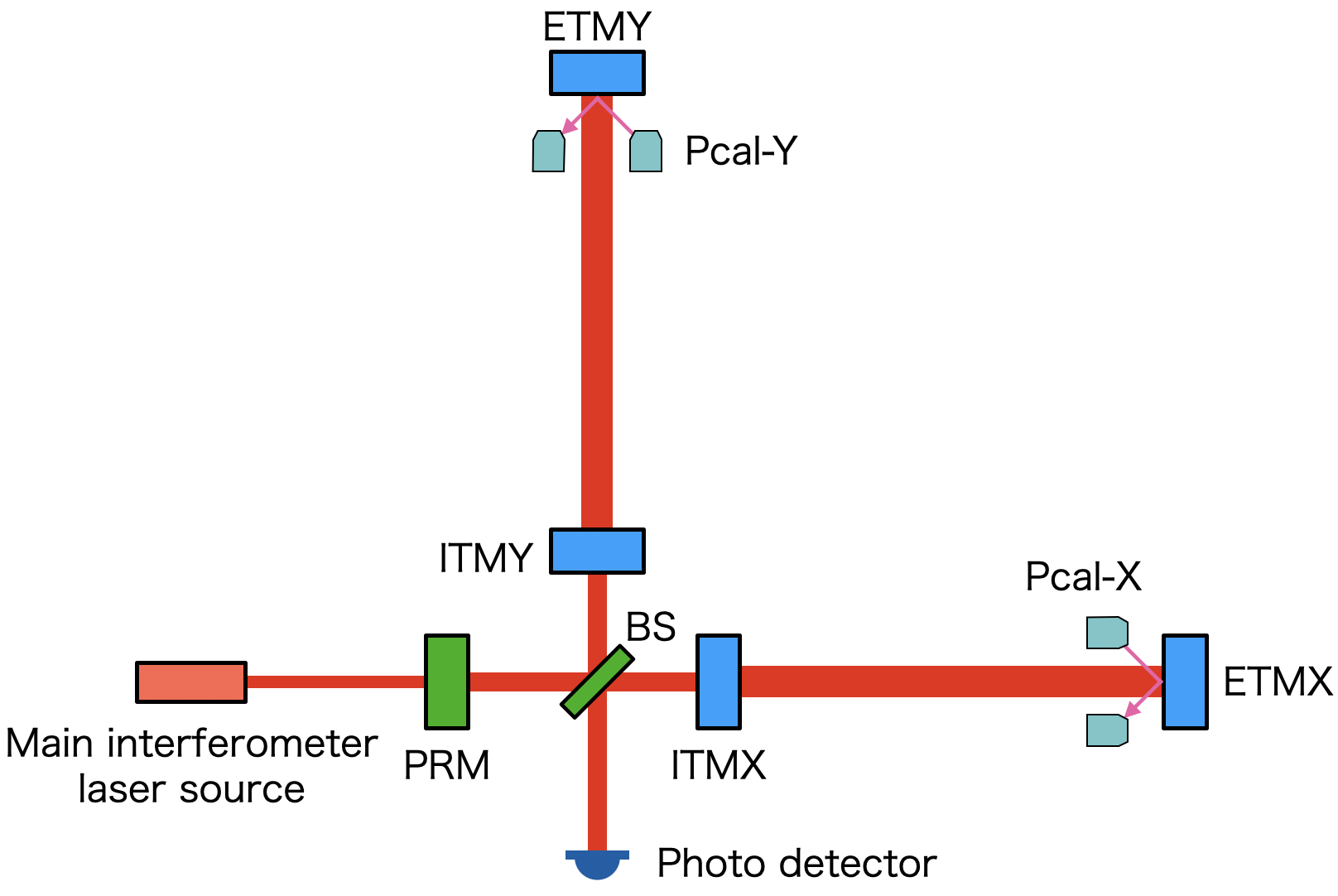}
        \caption{KAGRA interferometer and Pcal configuration during O4a. The PRFPMI consisted of the X- and Y-arm Fabry--Pérot cavities formed by ITMX--ETMX and ITMY--ETMY, respectively, the beam splitter (BS), and the power-recycling mirror (PRM). Pcal-X and Pcal-Y are installed near ETMX and ETMY, respectively, to inject calibrated reference signals into the interferometer via radiation pressure.}
        \label{fig:PRFPMI}
    \end{figure}

 KAGRA has two Pcal systems, located at the X- and Y-arm end stations and referred to as Pcal-X and Pcal-Y, respectively~\cite{10.1063/5.0147888,Chen_2025}.
    During O4a, Pcal-X, which applies reference signals to ETMX, was used as the primary calibration reference, while Pcal-Y at ETMY served as a stand-by system~\cite{Chen_2025}.
    A schematic overview of the interferometer configuration, including the locations of the two Pcal systems, is shown in Fig.~\ref{fig:PRFPMI}.
    
    \par In the LIGO--Virgo--KAGRA detectors, the laser power of the Pcal beams is measured using integrating-sphere power sensors, whose absolute calibration is traceable to national standards \cite{Bhattacharjee_2021,10.1063/1.4967303,Estevez_2021}.
    The uncertainty of the laser power measurement is the dominant contribution to the overall Pcal uncertainty.
    A detailed evaluation of the KAGRA Pcal performance during O4a was reported in Ref.~\cite{Chen_2025}, where the total uncertainty was estimated by propagating the uncertainties of individual parameters in the calibration model.

    \par However, in the O4a analysis of the KAGRA Pcal system, the scatter of repeated calibration measurements obtained under nominally identical conditions was larger than expected from the uncertainties assigned to the individual measurements.
    In Ref.~\cite{Chen_2025}, this excess scatter was empirically reflected in the reported calibration uncertainties, while its physical origin was not identified.
    Identifying the physical sources of this excess scatter is important for establishing a more physically motivated uncertainty model and for improving the reliability of the Pcal system.
    
    \par In this work, we investigate whether variations in the incident-beam conditions of the integrating-sphere power sensors can account for this excess scatter.
    We experimentally characterize the dependence of the integrating-sphere response on incidence angle, polarization, and beam position, and explicitly incorporate the corresponding systematic uncertainties into the uncertainty model for the O4a Pcal-X calibration measurements.

\section{Background: Pcal power calibration and observed excess scatter}
\label{sec:background}

    \subsection{KAGRA Pcal}
    \label{subsec:KAGRA_Pcal}
        \par
\begin{figure}
            \centering
            \includegraphics[width=0.75\linewidth]{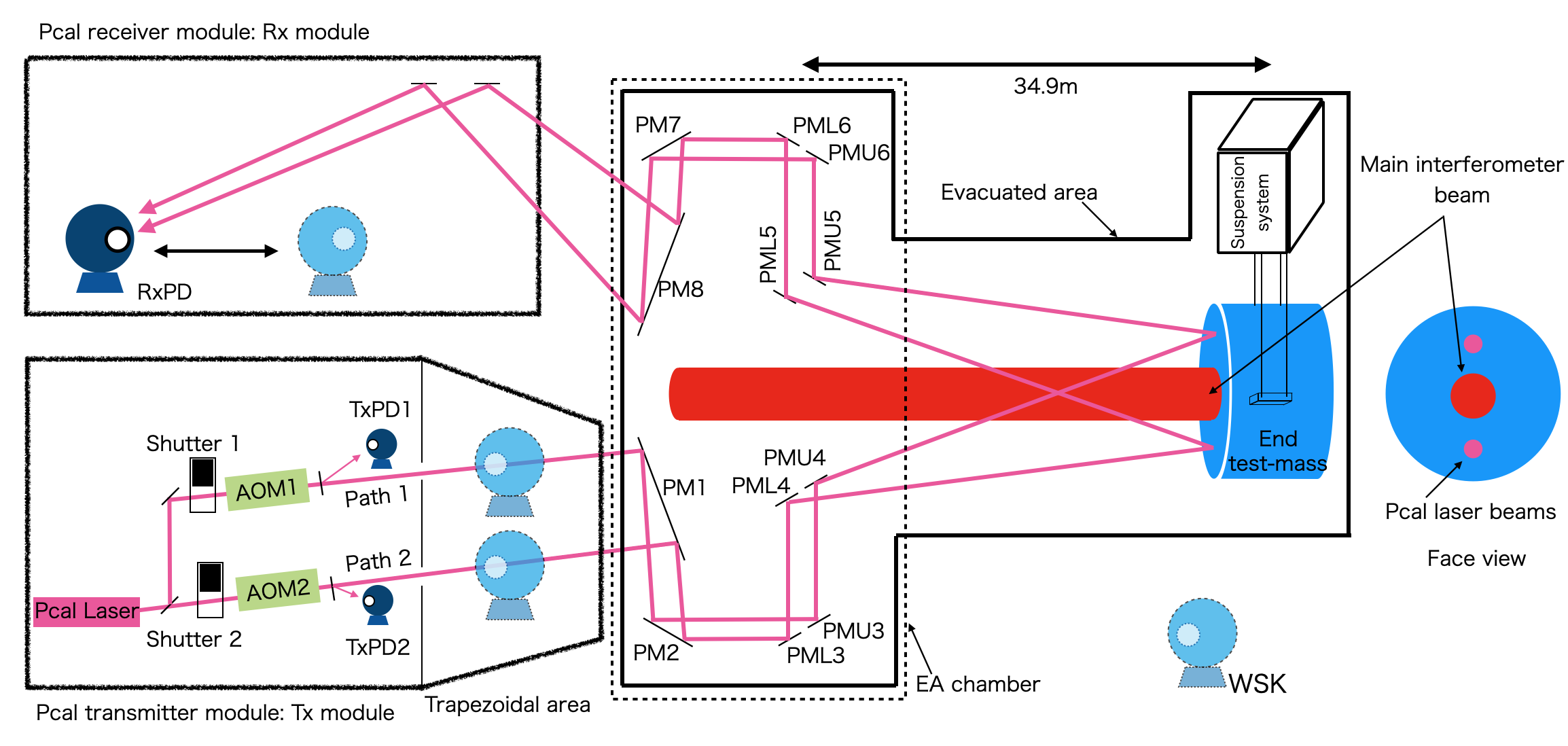}
            \caption{
                Schematic overview of the KAGRA Pcal system.
                Two Pcal beams generated in the Tx module are directed to the ETM and are subsequently collected by the Rx module.
                The laser powers are monitored by integrating-sphere power sensors in the Tx and Rx modules.
                The working standard of KAGRA (WSK) is temporarily placed at the Tx and Rx locations to calibrate these power sensors.
            }
            \label{fig:Pcal}
        \end{figure}

 A schematic overview of the KAGRA Pcal system is shown in Fig.~\ref{fig:Pcal}.
        Each Pcal system consists primarily of a transmitter (Tx) module and a receiver (Rx) module.
        The Tx module generates two Pcal beams, referred to as paths 1 and 2, which are directed to the ETM, and the reflected beams are collected by the Rx module.
        Integrating-sphere power sensors installed in the Tx and Rx modules are used to monitor the laser power; the two sensors in the Tx module are referred to as TxPD1 and TxPD2 (collectively, TxPDs), while the sensor in the Rx module is referred to as RxPD.
        The displacement induced by the Pcal system can be described in the frequency domain as
        \begin{equation}
            \label{eq:Pcal}
            x(f) = -\frac{2P(f)\cos\theta}{c M (2\pi f)^2}
            \left[ 1 + (\vec{a}\cdot\vec{b}) \frac{M}{I} \right],
        \end{equation}
        where $P(f)$ is the total laser power reflected from the ETM, $\theta$ is the incidence angle of the Pcal beams, $c$ is the speed of light, $M$ is the mass of the ETM, and $I$ is its moment of inertia~\cite{Goetz_2009,Goetz_2010,10.1063/1.4967303,Chen_2025}.
        The vector $\vec{a}$ is the power-weighted displacement vector of the Pcal beams from the center of the mirror surface, while $\vec{b}$ is the displacement vector of the interferometer beam from the center of the ETM surface.

        \par A detailed description of the KAGRA Pcal system and the evaluation of these parameters during O4a is given in Ref.~\cite{Chen_2025}.
        In the present work, we focus on the calibration of the Pcal laser power $P$, because its uncertainty was the dominant contribution to the overall Pcal uncertainty during O4a.
        The other parameters in Eq.~\ref{eq:Pcal}, including the beam-position-dependent rotational correction, are taken from Ref.~\cite{Chen_2025} unless otherwise stated.

    \subsection{Pcal power estimation}
    \label{subsec:Pcal_power_estimation}

        \par
\begin{figure}
            \centering
            \includegraphics[width=0.75\linewidth]{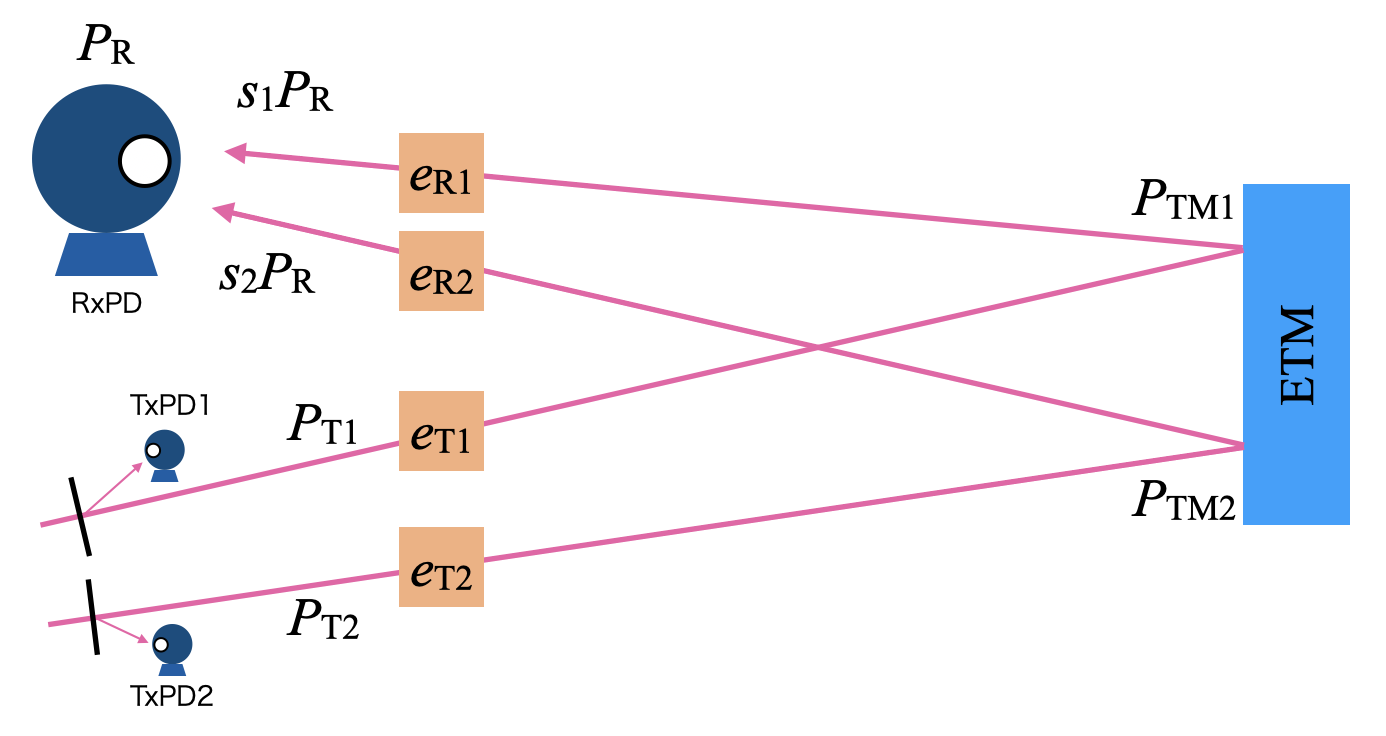}
            \caption{
                Schematic of the laser-power estimation at the ETM.
                The quantities $e_{Tj}$ and $e_{Rj}$ represent the optical efficiencies before and after reflection from the ETM, respectively, and $s_j$ denotes the fractional contribution of each path to the power measured by the RxPD.
            }
            \label{fig:OE}
        \end{figure}

 The laser power reflected from the ETM cannot be measured directly. Instead, it is inferred from the powers measured by integrating-sphere power sensors installed in the Tx and Rx modules. The Tx module monitors the outgoing laser powers for the two paths, denoted as $P_{T1}$ and $P_{T2}$, while the Rx module measures the total reflected power $P_R$, as illustrated in Fig.~\ref{fig:OE}.

        \par The optical efficiencies from the Tx module to the ETM and from the ETM to the Rx module are denoted as $e_{Tj}$ and $e_{Rj}$, respectively, with
        \begin{equation}
            e_j = e_{Tj} e_{Rj},
        \end{equation}
        where $j=1,2$.
        Following Ref.~\cite{Chen_2025}, the efficiencies before and after the ETM are assumed to be equal, $e_{Tj}=e_{Rj}=\sqrt{e_j}$.
        Since the RxPD measures the combined power from the two paths, separation ratios $s_1$ and $s_2$ describe their fractional contributions.

        \par Using these quantities, the total Pcal laser power $P$ reflected from the ETM is estimated by averaging the Tx- and Rx-side estimates:
        \begin{align}
            P &= P_{\mathrm{TM1}} + P_{\mathrm{TM2}} \\
              &= \frac{
                    P_{T1}e_{T1}
                    + P_{T2}e_{T2}
                    + s_1e_{R1}^{-1}P_R
                    + s_2e_{R2}^{-1}P_R
                 }{2}.
            \label{eq:Pcal_power_estimation}
        \end{align}
        The detailed derivation of this expression and the evaluation of the separation ratios are described in Ref.~\cite{Chen_2025}.

        \par
\begin{figure}
            \centering
            \includegraphics[width=0.75\linewidth]{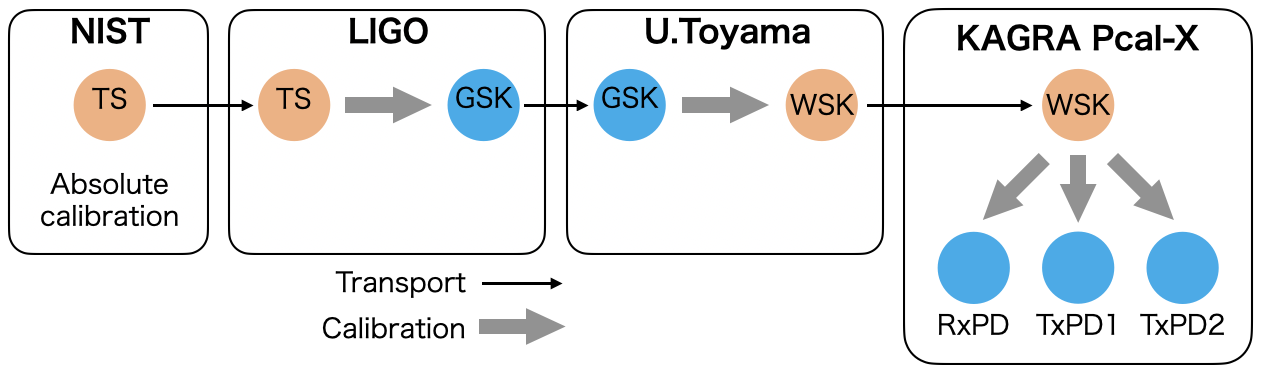}
            \caption{
                Transfer of the absolute laser-power calibration to the KAGRA Pcal power sensors.
                The calibration is transferred from a NIST-calibrated transfer standard to GSK, from GSK to WSK at the University of Toyama, and from WSK to the Pcal power sensors at the KAGRA end stations.
            }
            \label{fig:calibration_transportation}
        \end{figure}

 To convert the output voltages of the Pcal power sensors into laser power, their responsivities $\rho$ [V/W] must be calibrated. The absolute calibration is transferred through the calibration chain shown in Fig.~\ref{fig:calibration_transportation}~\cite{Bhattacharjee_2021,Chen_2025}.
        A transfer standard calibrated by the National Institute of Standards and Technology (NIST) was used at LIGO to calibrate the KAGRA gold standard (GSK).
        Following this calibration, GSK was transported from LIGO in the United States to Japan and is maintained at the University of Toyama (U. Toyama), where it is used to calibrate the KAGRA working standard (WSK).
        WSK is then transported to the KAGRA end stations and used to calibrate the RxPDs and TxPDs.

        \par The responsivity ratio $\alpha_{A/B}$ is defined as the ratio of the outputs of two power sensors $A$ and $B$ when they measure the same laser power. The responsivities of the Pcal power sensors can therefore be written as
        \begin{align}
            \rho_{\mathrm{Rx}}
                &= \rho_{\mathrm{GSK}}
                \alpha_{\mathrm{WSK/GSK}}
                \alpha_{\mathrm{RxPD/WSK}}, \\
            \rho_{\mathrm{TxPD1}}
                &= \rho_{\mathrm{GSK}}
                \alpha_{\mathrm{WSK/GSK}}
                \alpha_{\mathrm{TxPD1/WSK}}, \\
            \rho_{\mathrm{TxPD2}}
                &= \rho_{\mathrm{GSK}}
                \alpha_{\mathrm{WSK/GSK}}
                \alpha_{\mathrm{TxPD2/WSK}}.
        \end{align}
        The detailed calibration procedure and the absolute responsivity of GSK are given in Ref.~\cite{Chen_2025}.

        \par
\begin{figure}
            \centering
            \includegraphics[width=0.75\linewidth]{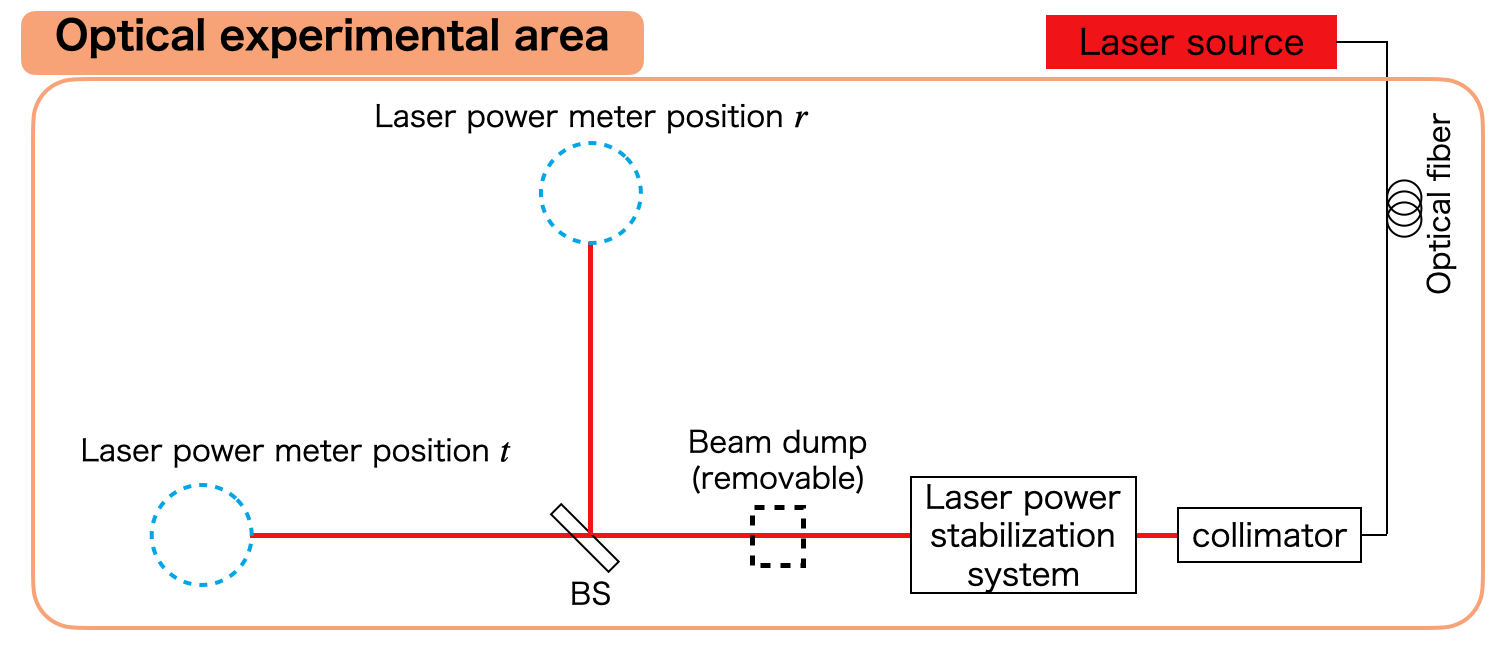}
            \caption{
                Experimental setup for the comparison calibration of GSK and WSK at the U. Toyama.
                The two integrating-sphere power sensors are alternately placed at the reflected and transmitted ports of the beam splitter.
            }
            \label{fig:WSK_CAL_setup}
        \end{figure}

 The GSK--WSK comparison measurement at the U. Toyama is illustrated in Fig.~\ref{fig:WSK_CAL_setup}. A power-stabilized laser is split by a beam splitter and injected into GSK and WSK. To reduce the influence of the beam-splitter reflectance and transmittance, GSK and WSK are alternately placed at the reflected ($r$) and transmitted ($t$) ports. This sequence is repeated five times during each calibration session. Following Ref.~\cite{Chen_2025}, the responsivity ratio obtained from the $i$-th comparison is calculated as
        \begin{align}
            \alpha_{i,\mathrm{WSK/GSK}}
            &=
            \sqrt{
            \left\langle
            \frac{
                \boldsymbol{V}_{i,r\mathrm{WSK}}
                - \langle \boldsymbol{V}_{i,r\mathrm{WSK},d} \rangle
            }{
                \boldsymbol{V}_{i,t\mathrm{GSK}}
                - \langle \boldsymbol{V}_{i,t\mathrm{GSK},d} \rangle
            }
            \right\rangle
            \left\langle
            \frac{
                \boldsymbol{V}_{i,t\mathrm{WSK}}
                - \langle \boldsymbol{V}_{i,t\mathrm{WSK},d} \rangle
            }{
                \boldsymbol{V}_{i,r\mathrm{GSK}}
                - \langle \boldsymbol{V}_{i,r\mathrm{GSK},d} \rangle
            }
            \right\rangle
            },
            \label{eq:alpha_i}
            \\
            \alpha_{\mathrm{WSK/GSK}}
            &=
            \left\langle
                \left[
                    \alpha_{1,\mathrm{WSK/GSK}},
                    \ldots,
                    \alpha_{5,\mathrm{WSK/GSK}}
                \right]
            \right\rangle.
            \label{eq:alpha_total}
        \end{align}
        Here, $V^{i,p}_{S}$ denotes the output voltage of power sensor $S$ at beam-splitter port $p$ ($p=r$ or $t$) in the $i$th comparison.
        The subscript $d$ denotes the corresponding dark-offset data, and $\langle \cdots \rangle$ denotes an average as defined in Appendix~\ref{app:average}.

        \par At the KAGRA end station, WSK is used as the reference sensor to calibrate the RxPD and TxPDs. The calibration measurements consist of measurements with the installed Pcal sensors, measurements with WSK placed at the Rx location, and measurements with WSK placed in the Tx beam paths. The detailed measurement sequence is described in Ref.~\cite{Chen_2025}. Here, we retain the expressions required for the uncertainty analysis in Sec.~\ref{sec:Uncertainty_modeling}.
        \par For the following equations, superscripts $A$, $B$, and $C$ denote measurements with the installed RxPD and TxPDs, with WSK placed at the Rx location, and with WSK placed at the Tx location, respectively. The quantities $V_{\rm Rx}$, $V_{\rm Tx,j}$, and $V_{\rm WSK,j}$ denote the output voltages of the RxPD, TxPD of path $j$, and WSK, respectively, and the subscript $d$ denotes the corresponding dark-offset voltage.
        \par The responsivity ratio between the RxPD and WSK for path $j$ is calculated as
        \begin{align}
            \alpha_{\mathrm{RxPD/WSK},j}
                &= \frac{
                    \langle \boldsymbol{R}_{\mathrm{Rx},j} \rangle
                }{
                    \langle \boldsymbol{R}_{\mathrm{WSK},j} \rangle
                },
                \label{eq:alpha_rxpd_wsk_1}
                \\
            \boldsymbol{R}_{\mathrm{Rx},j}
                &=
                \frac{
                    \boldsymbol{V}_{A,\mathrm{Rx},j}
                    - V_{\mathrm{RxPD},d}
                }{
                    \boldsymbol{V}_{A,\mathrm{Tx},j}
                    - V_{\mathrm{Tx},j,d}
                },
                \label{eq:alpha_rxpd_wsk_2}
                \\
            \boldsymbol{R}_{\mathrm{WSK},j}
                &=
                \frac{
                    \kappa_{\mathrm{WSK}}
                    \left(
                        \boldsymbol{V}_{B,\mathrm{WSK},j}
                        - V_{\mathrm{WSK},d}
                    \right)
                }{
                    \boldsymbol{V}_{B,\mathrm{Tx},j}
                    - V_{\mathrm{Tx},j,d}
                }.
                \label{eq:alpha_rxpd_wsk_3}
        \end{align}
        Here, $\kappa_{\mathrm{WSK}}$ corrects the difference in the WSK voltage readout between the U. Toyama and the KAGRA end station.

        \par The responsivity ratio between TxPD$j$ and WSK is
        \begin{equation}
            \alpha_{\mathrm{TxPD}j/\mathrm{WSK}}
            =
            \left\langle
                \frac{
                    \boldsymbol{V}_{C,\mathrm{Tx},j}
                    - V_{\mathrm{Tx},j,d}
                }{
                    \kappa_{\mathrm{WSK}}
                    \left(
                        \boldsymbol{V}_{C,\mathrm{WSK},j}
                        - V_{\mathrm{WSK},d}
                    \right)
                }
            \right\rangle .
            \label{eq:alpha_txpd_wsk}
        \end{equation}

        \par The overall optical efficiency $e_j$ is obtained from measurements with WSK at the Rx and Tx locations:
        \begin{align}
            e_j
                &=
                \frac{
                    \langle \boldsymbol{R}_{\mathrm{WSK},\mathrm{Rx},j} \rangle
                }{
                    \langle \boldsymbol{R}_{\mathrm{WSK},\mathrm{Tx},j} \rangle
                },
                \label{eq:e_j_1}
                \\
            \boldsymbol{R}_{\mathrm{WSK},\mathrm{Rx},j}
                &=
                \frac{
                    \boldsymbol{V}_{B,\mathrm{WSK},j}
                    - V_{\mathrm{WSK},d}
                }{
                    \boldsymbol{V}_{B,\mathrm{Tx},j}
                    - V_{\mathrm{Tx},j,d}
                },
                \label{eq:e_j_2}
                \\
            \boldsymbol{R}_{\mathrm{WSK},\mathrm{Tx},j}
                &=
                \frac{
                    \boldsymbol{V}_{C,\mathrm{WSK},j}
                    - V_{\mathrm{WSK},d}
                }{
                    \boldsymbol{V}_{C,\mathrm{Tx},j}
                    - V_{\mathrm{Tx},j,d}
                }.
                \label{eq:e_j_3}
        \end{align}

        \par The separation ratios $s_j$ are also determined from the end-station calibration measurements and are required for the laser-power estimation in Eq.~\ref{eq:Pcal_power_estimation}. Since the present work does not modify their evaluation, the corresponding expressions are not repeated here and are given in Ref.~\cite{Chen_2025}.

        \par An important feature of these calibration procedures is that the integrating spheres are not handled in the same way. GSK and WSK were manually repositioned during the comparison measurement at the U. Toyama. At the KAGRA end station, WSK was manually placed at the Tx and Rx locations, and the RxPD was also repositioned during calibration, whereas the TxPDs remain fixed. Consequently, the reproducibility of the incident-beam conditions differed among the power sensors. This distinction becomes important when the systematic uncertainties associated with incidence angle, polarization, and beam position are introduced in Sec.~\ref{sec:Uncertainty_modeling}.

    \subsection{Observed excess scatter in O4a}
    \label{subsec:Observed_excess_scatter_in_O4a}

        \par
\begin{figure}[htbp]
            \centering
            
            \begin{subfigure}{0.48\linewidth}
                \centering
                \includegraphics[width=\linewidth]{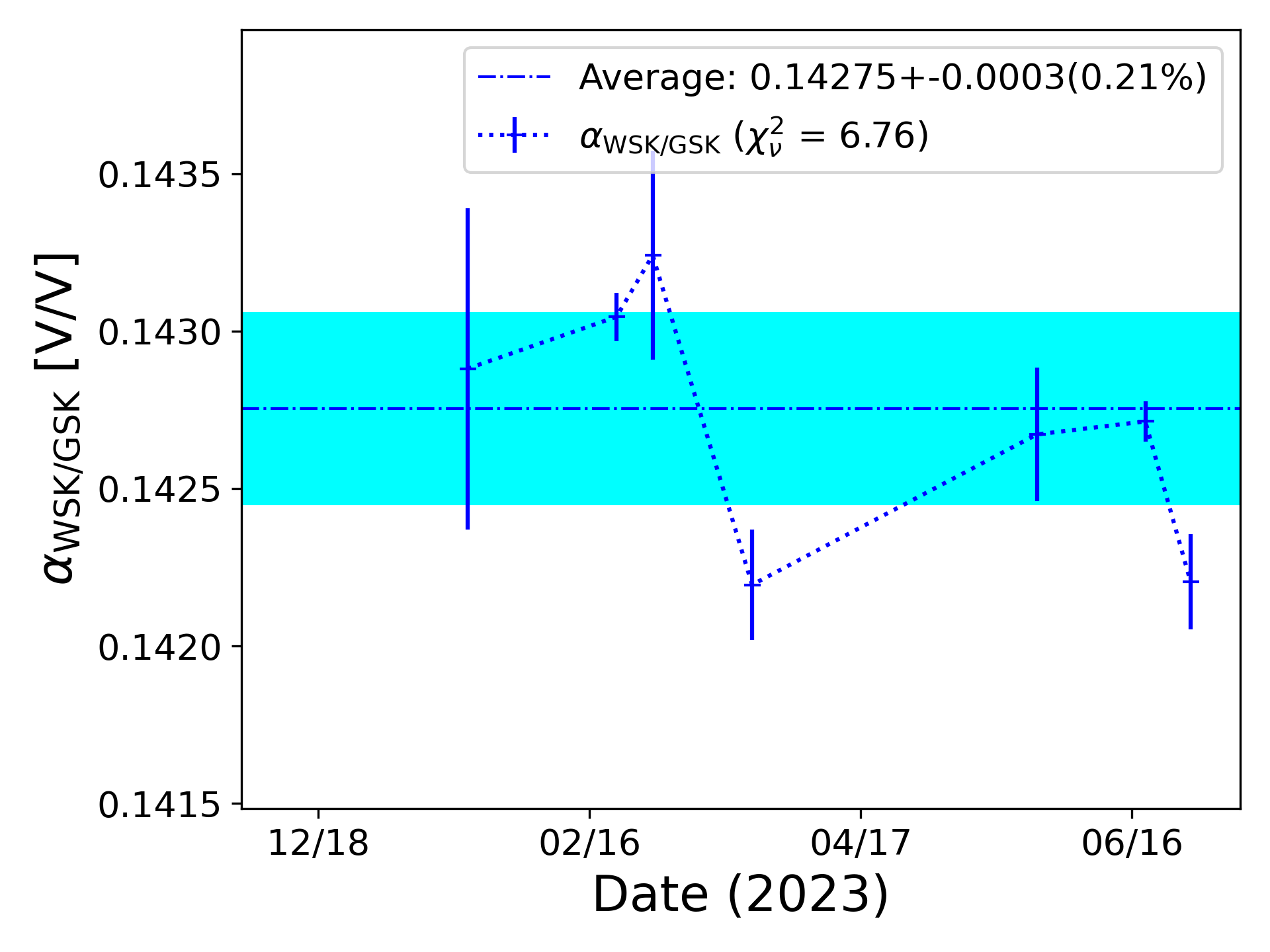}
                %\caption{$\alpha_{\mathrm{WSK/GSK}}$}
            \end{subfigure}
            \hfill
            \begin{subfigure}{0.48\linewidth}
                \centering
                \includegraphics[width=\linewidth]{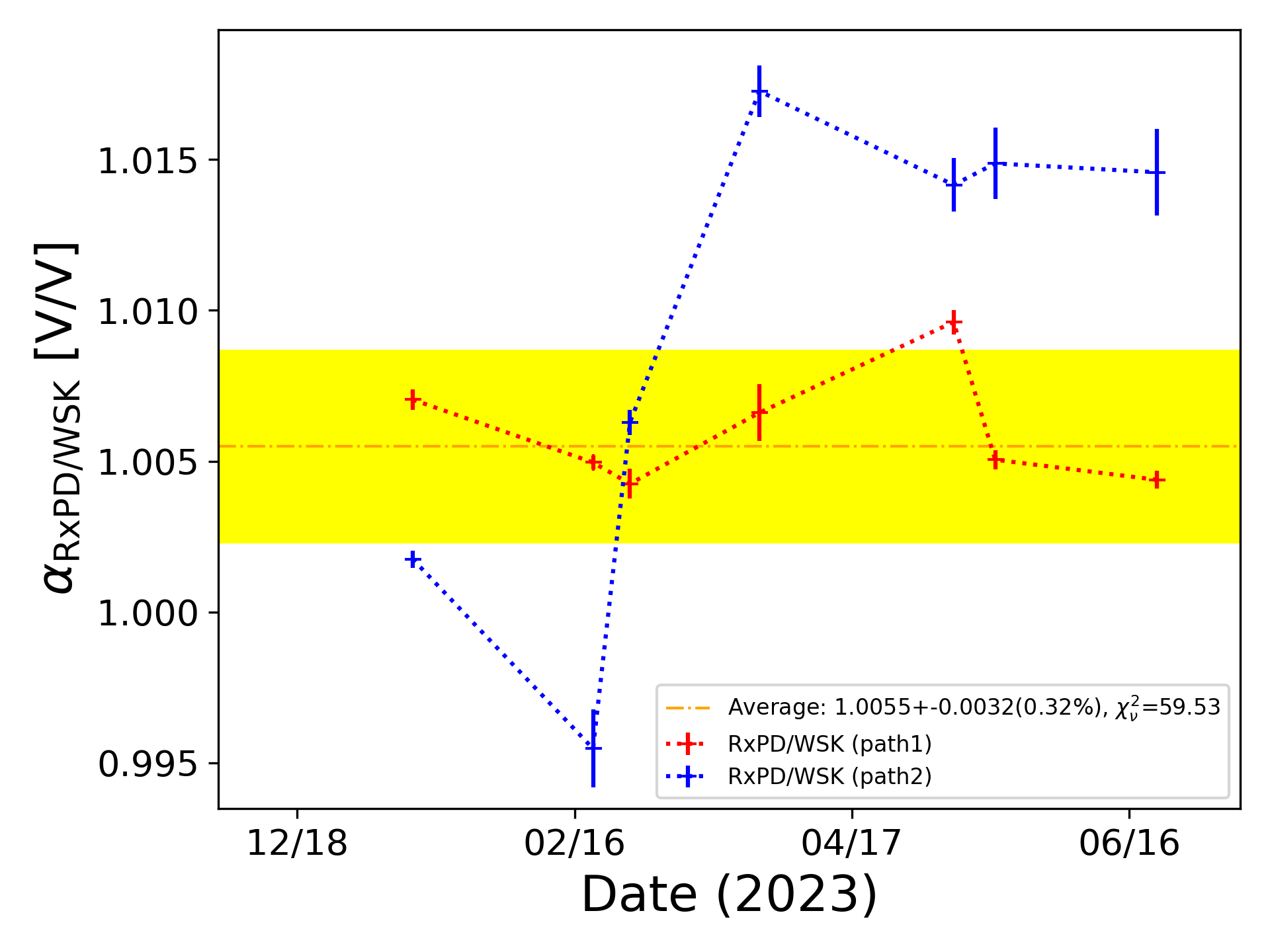}
                %\caption{$\alpha_{\mathrm{RxPD/WSK}}$}
            \end{subfigure}
            
            \vspace{0.3cm}
            
            \begin{subfigure}{0.48\linewidth}
                \centering
                \includegraphics[width=\linewidth]{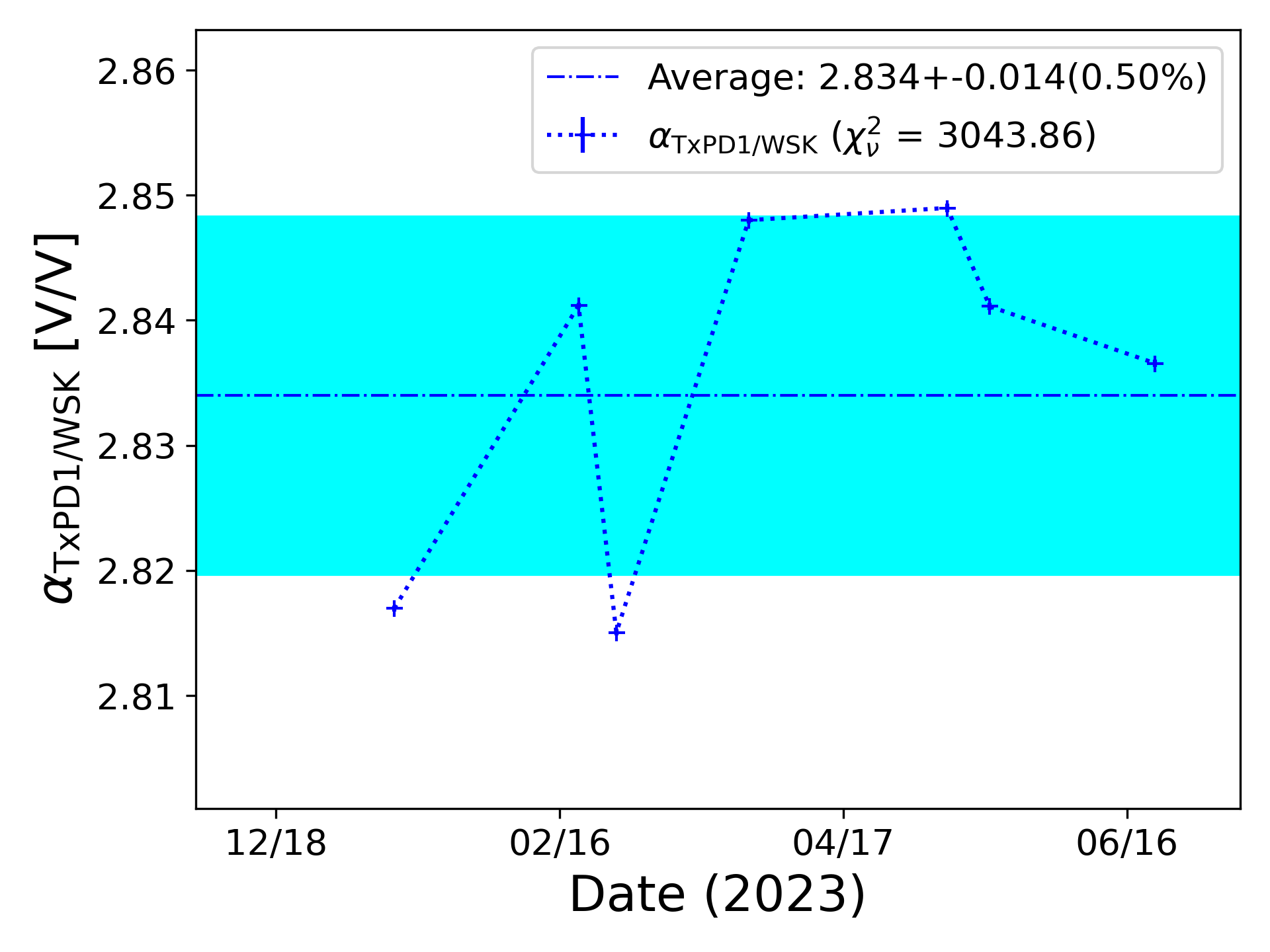}
                %\caption{$\alpha_{\mathrm{TxPD1/WSK}}$}
            \end{subfigure}
            \hfill
            \begin{subfigure}{0.48\linewidth}
                \centering
                \includegraphics[width=\linewidth]{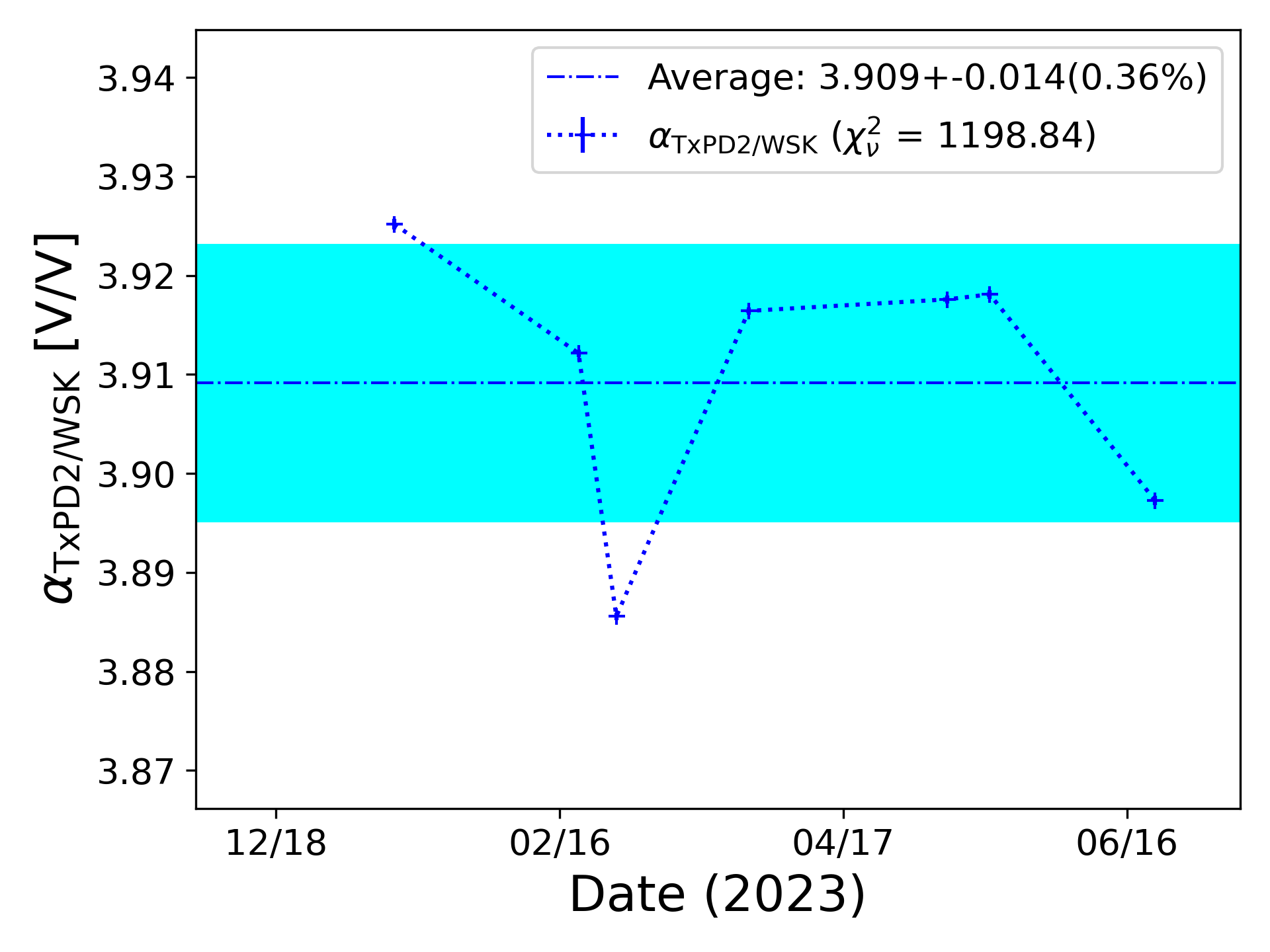}
                %\caption{$\alpha_{\mathrm{TxPD2/WSK}}$}
            \end{subfigure}

            \vspace{0.3cm}
            
            \begin{subfigure}{0.48\linewidth}
                \centering
                \includegraphics[width=\linewidth]{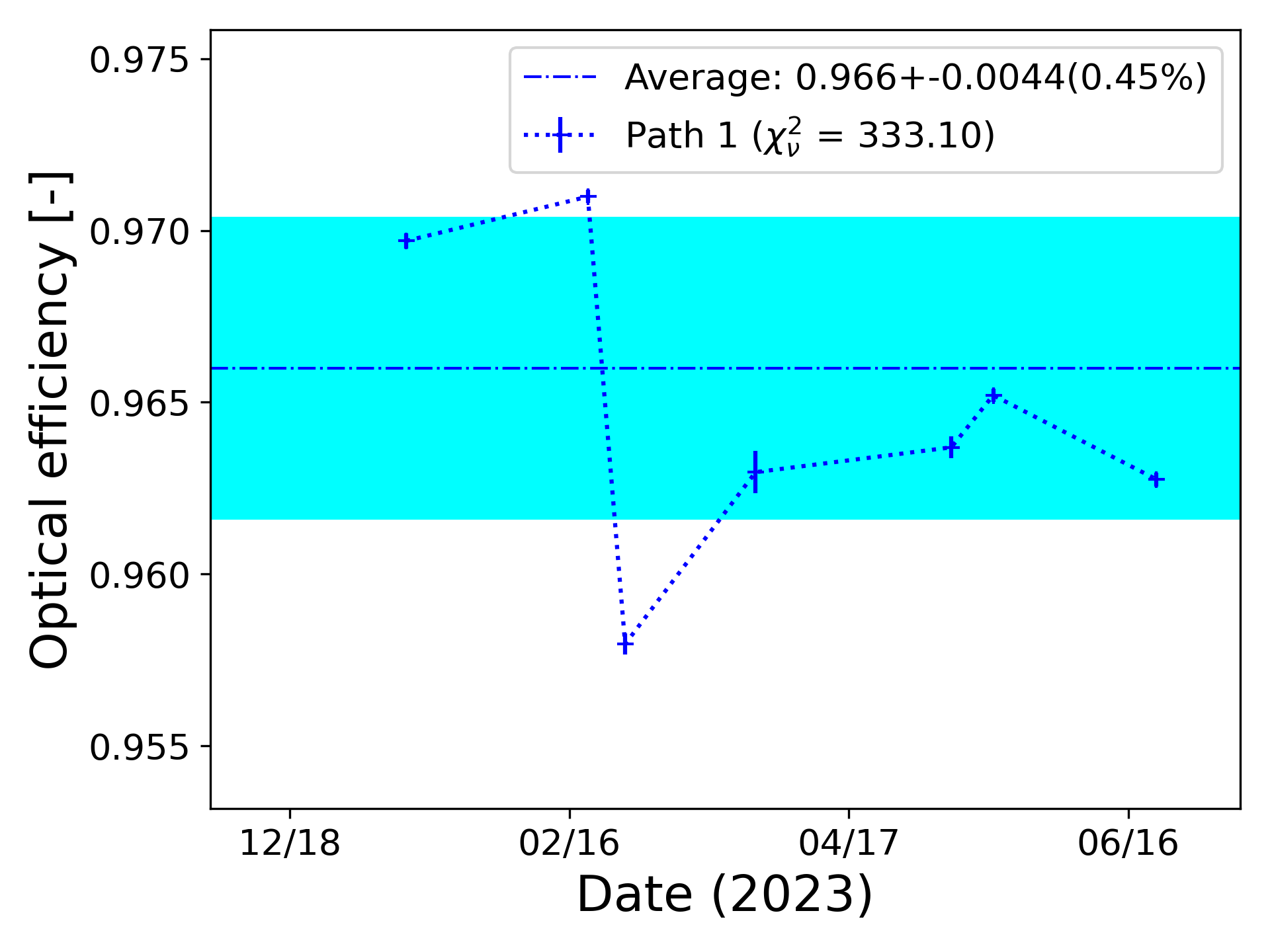}
                %\caption{$e_1$}
            \end{subfigure}
            \hfill
            \begin{subfigure}{0.48\linewidth}
                \centering
                \includegraphics[width=\linewidth]{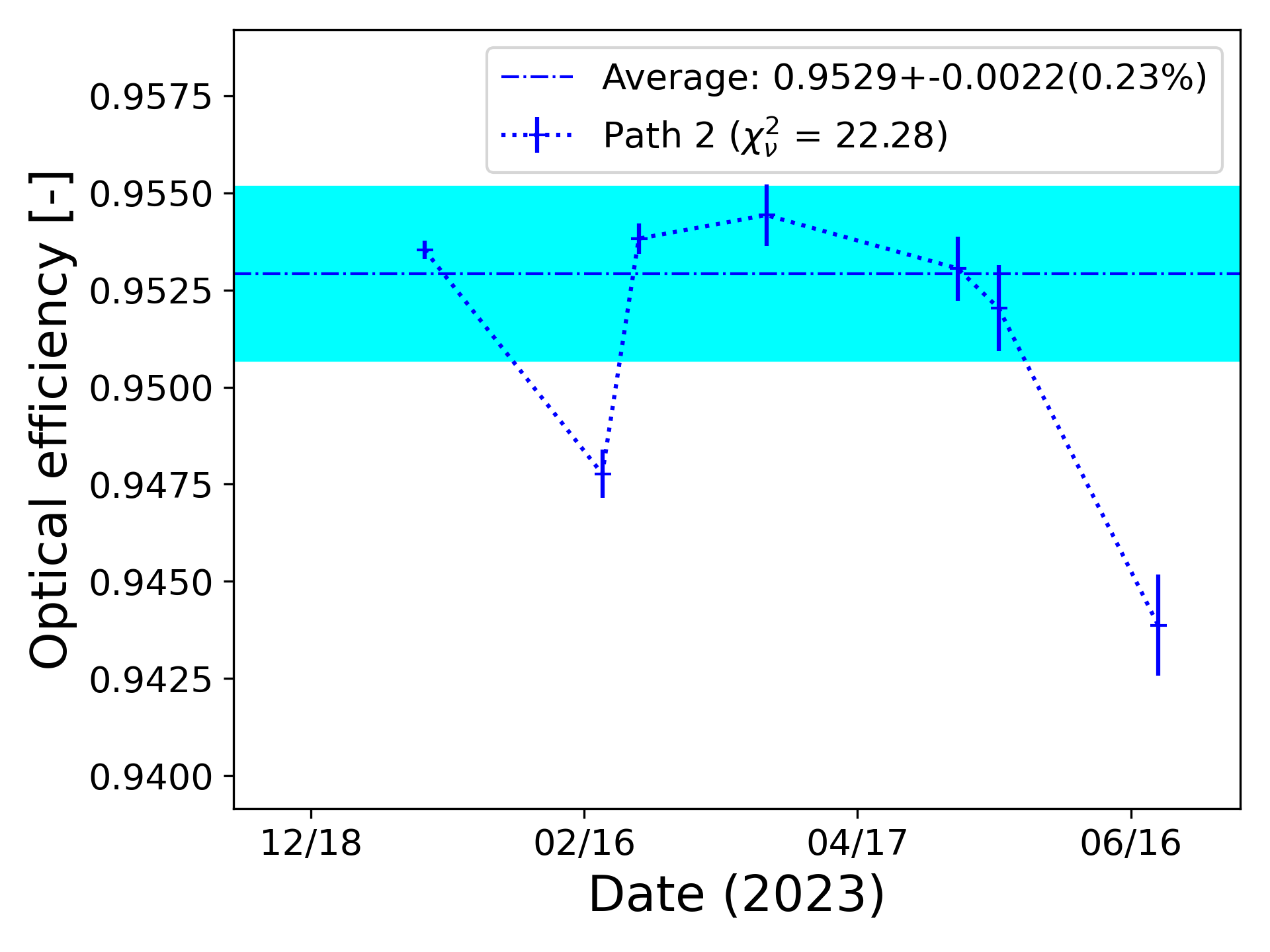}
                %\caption{$e_2$}
            \end{subfigure}
            
            \caption{
                Repeated calibration measurements relevant to the Pcal-X laser-power estimation during O4a.
                The O4a calibration data were previously reported in Ref.~\cite{Chen_2025} and are replotted here to illustrate the excess scatter relevant to the present study.
                The error bars represent the uncertainties assigned to the individual calibration measurements without the additional integrating-sphere systematic effects considered in this work.
                For several parameters, the observed scatter is substantially larger than the individual uncertainties, resulting in reduced chi-square values significantly greater than unity.
            }
            \label{fig:NOSYS}
        \end{figure}

 The Pcal-X power-sensor calibration parameters measured before and during O4a were reported in Ref.~\cite{Chen_2025}.
        Figure~\ref{fig:NOSYS} shows the responsivity ratios
        $\alpha_{\rm WSK/GSK}$, $\alpha_{\rm RxPD/WSK}$,
        $\alpha_{\rm TxPD1/WSK}$, and $\alpha_{\rm TxPD2/WSK}$,
        together with the optical efficiencies $e_1$ and $e_2$.
        
        \par For several parameters, the scatter among repeated calibration measurements is substantially larger than the uncertainties assigned to the individual measurements.
        This discrepancy is reflected in reduced chi-square values $\chi^2_{\rm \nu}$ significantly greater than unity.
        In particular, the discrepancy is pronounced for several parameters measured at the KAGRA end station, indicating that the uncertainties assigned to the individual calibration measurements do not fully describe the observed measurement-to-measurement variation.
        
        \par The data shown in Fig.~\ref{fig:NOSYS} therefore provide a basis for testing whether the variations in the incident-beam conditions described above can account for the observed excess scatter.
        In the following section, we experimentally characterize the response of the integrating spheres to these incident-beam conditions.

\section{Experimental characterization of integrating-sphere response}
\label{sec:Experimental}
    \par Because the beam size and laser power differ among the measurement locations, three models of integrating spheres are used in the KAGRA Pcal system, including those in the Tx and Rx modules at the end stations and those in the calibration system at U. Toyama for WSK calibration. The integrating spheres were manufactured by Labsphere Inc. Port-frame reducers are also used depending on the beam size. The model numbers, port-frame reducers, and actual input-port sizes are summarized in Table~\ref{table:IS}. The model names indicate the type of integrating sphere: LPM denotes a Laser Power Measurement sphere, whereas GPS denotes a General Purpose sphere. GSK is an LPM-type sphere, whereas WSK, RxPD, and the TxPDs are GPS-type spheres.
    In KAGRA, the Tx and Rx modules are located near the A-type chamber (EA chamber) at each end station, which houses the final Pcal steering mirrors approximately 34.9 m from the ETM.
    Owing to this long propagation distance, the Pcal laser beam has a diameter of approximately 12 mm, which is larger than those used in LIGO and Virgo. %cite LIGO, Virgoについて言及しているためここにあると良い
    Therefore, integrating spheres with larger input ports are used for WSK and RxPD to accommodate the beam. The other sensors use 1-inch input ports to minimize the effect of backscattered light from the integrating sphere.

    \begin{table}[htbp]
        \centering
        \caption{Specifications of the integrating spheres used in the KAGRA Pcal systems}
        \begin{tabular}{c c c c c}
        \hline
         & Model number & Inner diameter & Port-frame-reducer & Input port size \\
        \hline
        GSK   & 3P-LPM-040-SL & 4 inch & - & 1 inch \\
        WSK   & 3P-GPS-053-SL & 5.3 inch & PFR-FM/M-250-150-SF & 1.5 inch \\
        RxPD  & 3P-GPS-053-SL & 5.3 inch & PFR-FM/M-250-150-SF & 1.5 inch \\
        TxPD1 & 3P-GPS-020-SL & 2 inch & - & 1 inch \\
        TxPD2 & 3P-GPS-020-SL & 2 inch & - & 1 inch \\
        \hline
        \end{tabular}
        \label{table:IS}
    \end{table}

    \par Integrating spheres are generally expected to be relatively insensitive to variations in the incident-beam conditions~\cite{Labsphere_IS}. However, for the sub-percent-level precision required for Pcal laser-power calibration, even small residual dependences on the incident-beam conditions may become relevant. 
    %Therefore, in this study, we investigate three possible dependences of the integrating-sphere response: incidence angle, polarization, and beam position. %before short Sec3.1%
    Therefore, in this study, we consider three possible dependences of the integrating-sphere response: incidence angle, polarization, and beam position. The incidence-angle dependence was previously reported in Ref.~\cite{Fujii:2023+W} and is reanalyzed here for the uncertainty evaluation, whereas the polarization and beam-position dependences are newly characterized in the present study.

    \par
\begin{figure}
        \centering
        \begin{subfigure}{0.48\linewidth}
            \centering
            \includegraphics[width=\linewidth]{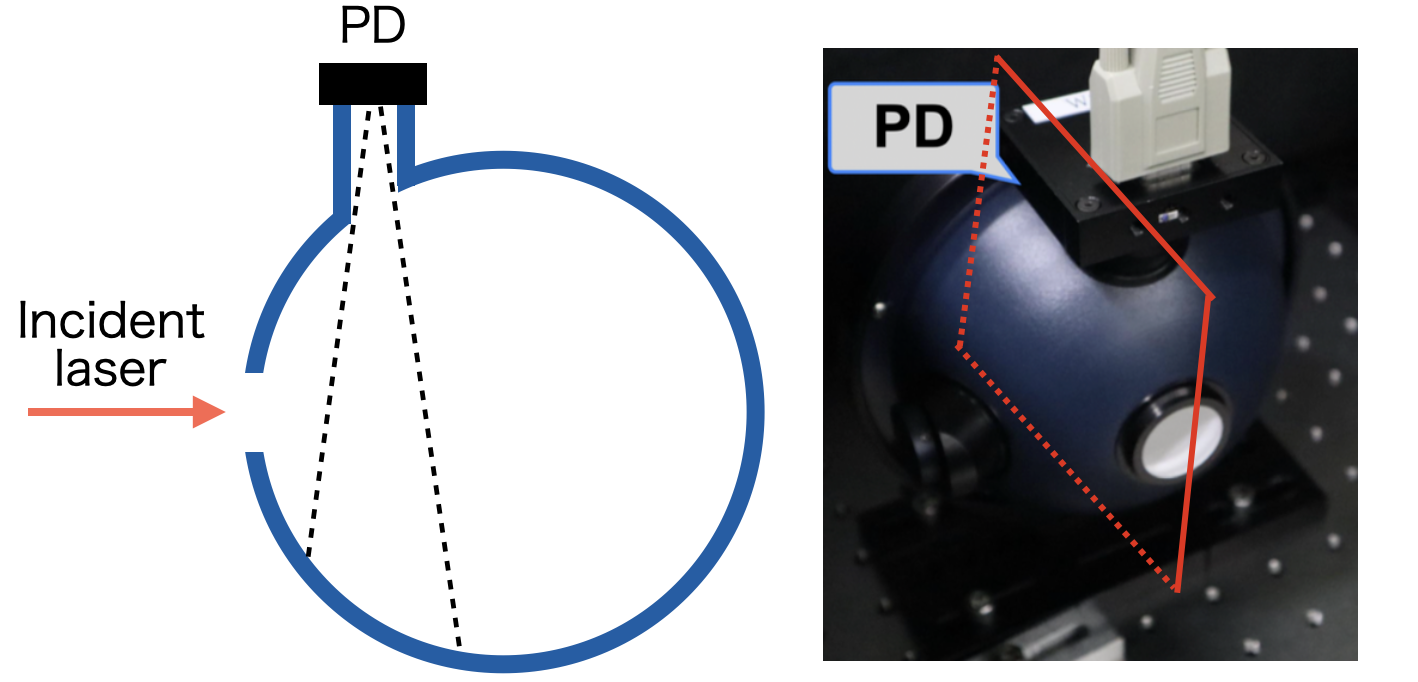}
            \caption{The GPS sphere}
        \end{subfigure}
        \hfill
        \begin{subfigure}{0.48\linewidth}
            \centering
            \includegraphics[width=\linewidth]{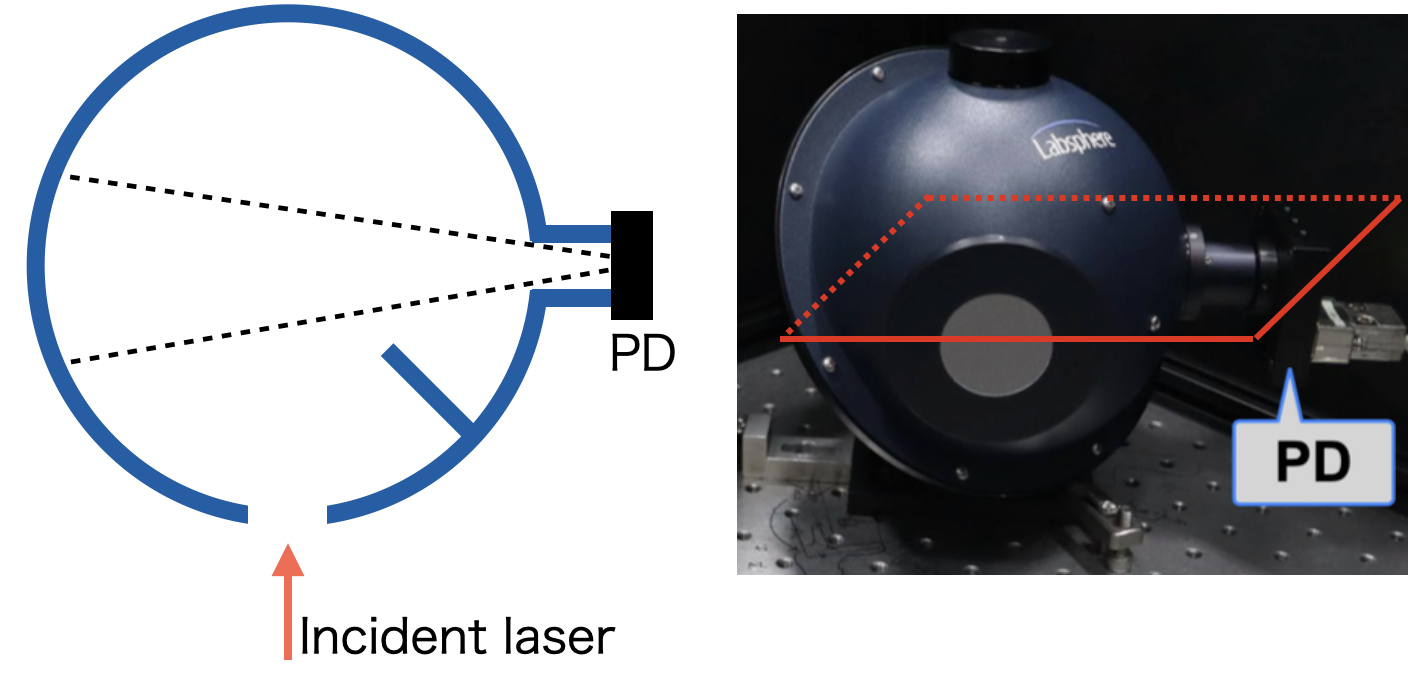}
            \caption{The LPM sphere}
        \end{subfigure}
        \caption{Two types of integrating spheres used in the KAGRA Pcal systems. For each panel, the schematic on the left shows the sphere in the plane indicated by the red lines in the photograph on the right. The GPS sphere contains a baffle between the input and detection ports, whereas the LPM sphere has no baffle.}
        \label{fig:IS_type}
    \end{figure}

 The dependence measurements were performed using two types of integrating spheres: an LPM-type sphere (3P-LPM-040-SL; hereafter, LPM sphere) and a GPS-type sphere (3P-GPS-053-SL; hereafter, GPS sphere). The LPM sphere used in this study has the same model as that used for GSK, whereas the GPS sphere is identical to WSK.
    The two types of integrating spheres have different internal structures, which may result in different dependences on the incident-beam conditions (Fig.~\ref{fig:IS_type}). The GPS sphere has a baffle between the input and detection ports, and the detection port is located 90° from the input port. In addition, the photodetector (PD) faces a region located a further 90° away. In contrast, the LPM sphere has no baffle, and the PD is located midway between the input port and the 90° position, so that it faces an intermediate region rather than the region at 90°.

\subsection{Incidence-angle dependence}
\label{subsec:Incident_angle_dependence}
    \par The incidence-angle dependence of the LPM- and GPS-type integrating spheres was previously measured and reported in Ref.~\cite{Fujii:2023+W}.
    The GPS sphere used in that measurement was 3P-GPS-053-SL, identical to WSK, although a different model number was inadvertently reported in Ref.~\cite{Fujii:2023+W}.
    In the present study, we reanalyze the same data specifically to evaluate the systematic uncertainty associated with incidence-angle variations in the O4a Pcal calibration measurements.
    The original measurements were performed using the WSK calibration setup shown in Fig.~\ref{fig:WSK_CAL_setup}, with GSK fixed at position $t$ as a reference for fluctuations in the input laser power and the test integrating sphere placed at position $r$.
    The test integrating sphere was rotated around the center of its input port to vary the incidence angle while minimizing changes in the beam position on the input port.
    The incidence angle was varied from $-15^\circ$ to $15^\circ$ in $1^\circ$ increments.
    Normal incidence was defined as $\theta=0^\circ$, with clockwise and counterclockwise rotations viewed from above defined as negative and positive, respectively.
    Details of the experimental setup and measurement procedure are given in Ref.~\cite{Fujii:2023+W}.

    \par Let $\boldsymbol{V}_{\theta}$ and $\boldsymbol{V}_{{\rm ref},\theta}$ denote the output-voltage samples of the test integrating sphere and GSK, respectively, at incidence angle $\theta$, and let $\boldsymbol{V}_{\theta,{\rm d}}$ and $\boldsymbol{V}_{{\rm ref},\theta,{\rm d}}$ denote the corresponding dark-offset samples.
    At each incidence angle, the response normalized by the reference laser power was calculated as
    \begin{equation}
        R_{\mathrm{ang}}(\theta)
        =
        \left\langle
        \frac{
            \boldsymbol{V}_{\theta} - \langle \boldsymbol{V}_{\theta, {\rm d}} \rangle
        }{
            \boldsymbol{V}_{{\rm ref}, \theta} - \langle \boldsymbol{V}_{{\rm ref}, \theta, {\rm d}} \rangle
        }
        \right\rangle .
    \end{equation}
    The response relative to that at normal incidence was then calculated as
    \begin{equation}
        r_{\mathrm{ang}}(\theta)
        =
        \frac{
            R_{\mathrm{ang}}(\theta)
        }{
            R_{\mathrm{ang}}(0)
        }.
    \end{equation}

    \par
\begin{figure}
        \centering
        \begin{subfigure}{0.48\linewidth}
            \centering
            \includegraphics[width=\linewidth]{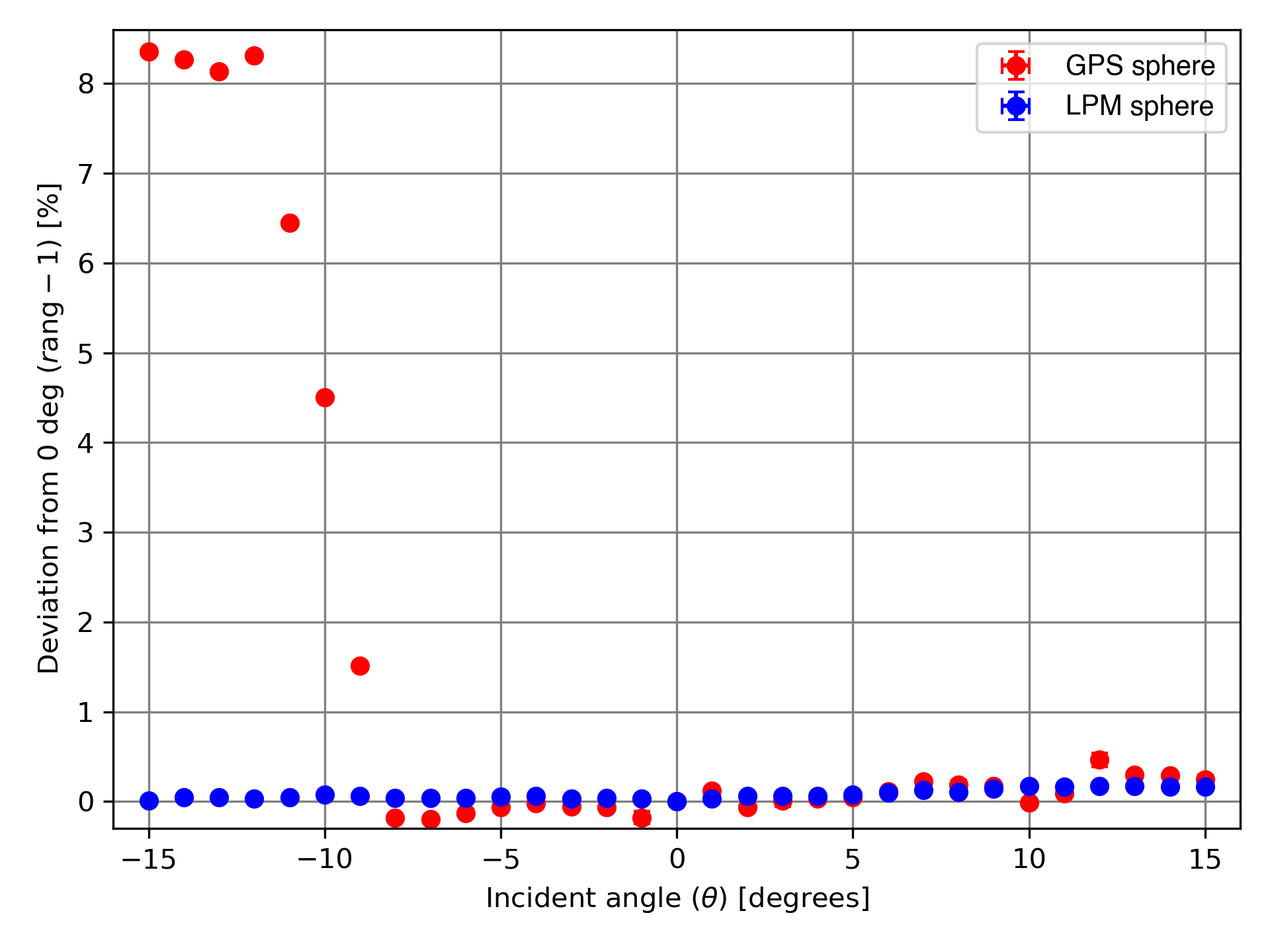}
            \caption{Full response variation.}
        \end{subfigure}
        \hfill
        \begin{subfigure}{0.48\linewidth}
            \centering
            \includegraphics[width=\linewidth]{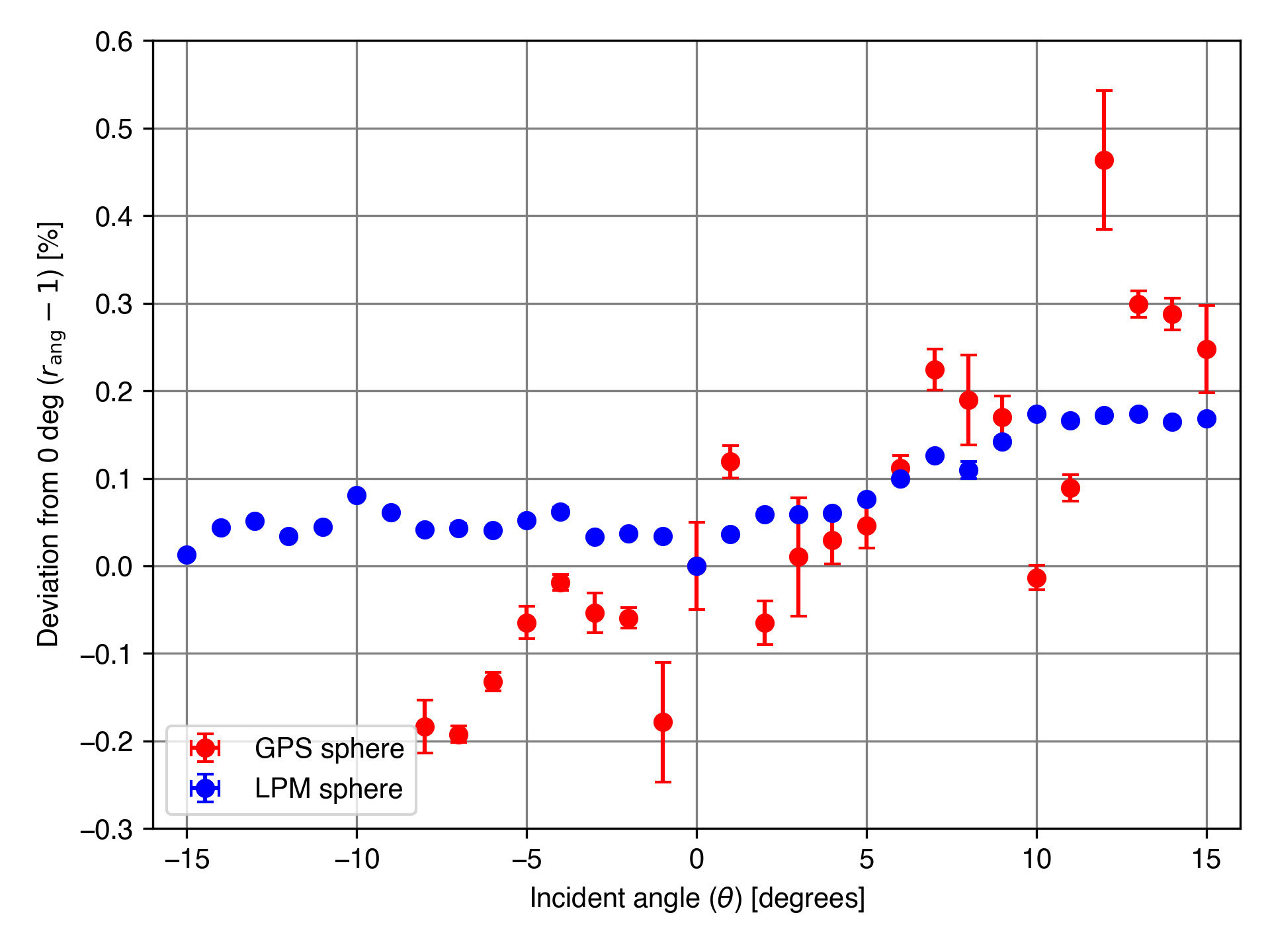}
            \caption{An enlarged view.}
        \end{subfigure}
        \caption{Results of the incidence-angle-dependence measurement reported in Ref.~\cite{Fujii:2023+W}. Panel (a) shows the full response variation over the measured angular range, while panel (b) shows an enlarged view with the vertical axis restricted to $-0.3\%$ to $0.8\%$ to highlight the smaller variations.}
        \label{fig:Angle_result}
    \end{figure}

 The results are shown in Fig.~\ref{fig:Angle_result}.
    An incidence-angle dependence was observed for both types of integrating spheres.
    For the LPM sphere, the response variation remained within $0.2\ \%$ over the measured range of $\pm 15^\circ$.
    In contrast, the GPS sphere showed a stronger angle dependence, with particularly large changes for angles below approximately $-9^\circ$.
    Near normal incidence, the GPS response was relatively stable but still showed a larger local variation than the LPM response.

    \par For the present uncertainty evaluation, the relevant quantity is the local response variation around the nominal incidence angle for each calibration configuration rather than the variation over the full measured angular range.
    Because the integrating-sphere response depends nonlinearly on the incidence angle, the corresponding uncertainty depends on the nominal incidence angle $\theta_0$.
    In the calibration measurements, the reproducibility of the incidence angle was assumed to be $\pm 2^\circ$.
    We therefore evaluated the local variation of the integrating-sphere response using the five measured points at
    $\theta=\theta_0-2^\circ,\ldots,\theta_0+2^\circ$.

    \par For these five points, the weighted mean response and weighted standard
    deviation are calculated using the procedure defined in
    Appendix~\ref{app:average}, with
    \begin{equation}
        \Theta(\theta_0)
        =
        \{
            \theta_0-2^\circ,
            \theta_0-1^\circ,
            \theta_0,
            \theta_0+1^\circ,
            \theta_0+2^\circ
        \}.
    \end{equation}
    The weighted standard deviation of the five response values is denoted by
    \begin{equation}
        s_{\mathrm{w}}(\theta_0)
        =
        s\left(
            R_{\mathrm{ang}}(\theta)
        \right)_{\theta\in\Theta(\theta_0)}.
    \end{equation}
    The relative uncertainty due to the incidence-angle variation is then defined as
    \begin{equation}
        u_{\mathrm{name,ang}|\theta_0}
        =
        \frac{
            s_{\mathrm{w}}(\theta_0)
        }{
            R_{\mathrm{ang}}(0)
        },
        \label{eq:angle_uncertainty}
    \end{equation}
    where ``name'' denotes either the LPM or GPS integrating sphere.

\subsection{Polarization dependence}
\label{subsec:Polarization_dependence}
    %\par The polarization-dependence measurement was performed using a setup similar to that used for the WSK calibration and the incidence-angle-dependence measurement, with the configuration at position “r” modified as shown in Fig.~\ref{fig:polarization_setup}. %before short Sec3.1%
    \par
\begin{figure}
        \centering
        \includegraphics[width=0.95\linewidth]{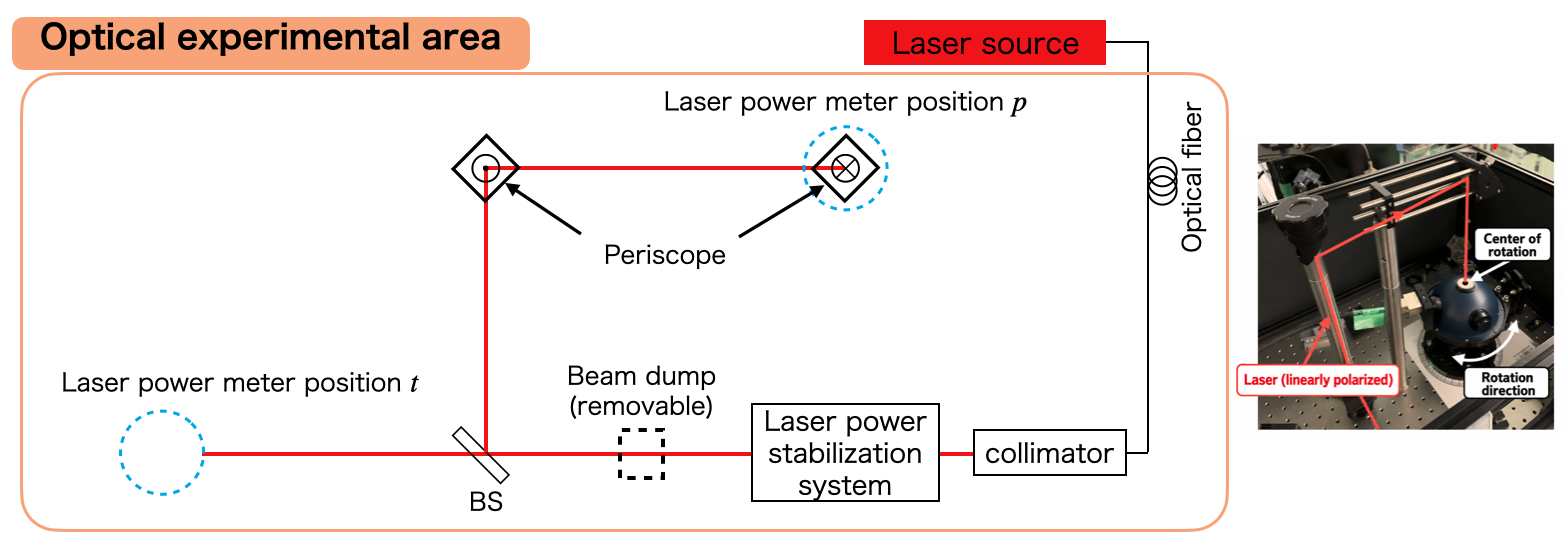}
        \caption{Experimental setup for the polarization-dependence measurement. The transmitted beam is directed to a reference power meter, while the reflected beam is directed through a periscope vertically downward onto the test integrating sphere. The sphere is rotated about the incident-beam axis to vary the polarization angle relative to the sphere.}
        \label{fig:polarization_setup}
    \end{figure}

 The polarization-dependence measurement was performed using a setup based on the WSK calibration setup shown in Fig.~\ref{fig:WSK_CAL_setup}, with the configuration at position $r$ modified as shown in Fig.~\ref{fig:polarization_setup}.
    Because a precision of approximately $ 0.1\ \%$ level or better was required, rotating the laser polarization using a half-wave plate was avoided.
    This is because rotating a half-wave plate can slightly shift the beam position on the optic, potentially introducing additional variations in optical loss.
    Instead, the test integrating sphere itself was rotated about the incident-beam axis, thereby changing the polarization angle relative to the sphere.
    The sphere was mounted on a rotation stage with its input port facing upward, and the rotation angle was determined using a protractor fixed below the stage.
    The laser beam was then directed vertically downward into the test integrating sphere using a periscope.
    
    \par GSK was placed at the $t$ position and used as the reference for the incident laser power, as in the incidence-angle-dependence measurement.
    The test integrating sphere was fixed at the $p$ position facing upward.
    The measurements were performed for both the LPM sphere and the GPS sphere.
    We define $\Theta$ as the rotation angle of the test integrating sphere about the incident-beam axis, corresponding to the polarization angle of the incident laser relative to the integrating sphere.
    The measurement procedure was as follows.

    \begin{enumerate}
        \item The polarization angle $\Theta$ was varied from $0^\circ$ to $230^\circ$
        in $10^\circ$ increments.
    
        \item At each polarization angle $\Theta$, the outputs from the test integrating sphere ($\boldsymbol{V}_{i,\Theta}$) and GSK ($\boldsymbol{V}_{i,\mathrm{ref},\Theta}$) were recorded simultaneously using an ADCMT 7352A Digital Multimeter.
        Each measurement consisted of 220 data points acquired over approximately 1~min.
    
        \item The measurement was repeated seven times at each polarization angle
        ($i = 1, \cdots, 7$).
    
        \item Beam-off data were also acquired for each polarization angle for both the test integrating sphere and GSK
        ($\boldsymbol{V}_{\Theta,\mathrm{d}}$ and
        $\boldsymbol{V}_{\mathrm{ref},\Theta,\mathrm{d}}$).
    \end{enumerate}
    
    \par The dark offsets were subtracted from the corresponding measurements of the test integrating sphere and GSK.
    The response normalized by the reference laser power was then calculated for each measurement and averaged as follows.
    \begin{align}
        R_{i, \mathrm{pol}}(\Theta)
        &=
        \left\langle
        \frac{
        \boldsymbol{V}_{i,\Theta} - \left\langle \boldsymbol{V}_{\Theta, \mathrm{d}} \right\rangle
        }{
        \boldsymbol{V}_{i,\mathrm{ref},\Theta} - \left\langle \boldsymbol{V}_{\mathrm{ref},\Theta,\mathrm{d}} \right\rangle
        }
        \right\rangle, \\
        R_{\mathrm{pol}}(\Theta)
        &=
        \left\langle
        R_{i,\mathrm{pol}}(\Theta)
        \right\rangle_{i=1,\cdots,7}.
    \end{align}

    The mean response over all measured polarization angles is defined as
    \begin{equation}
        \overline{R}_{\mathrm{pol}}
        =
        \left\langle
        R_{\mathrm{pol}}(\Theta)
        \right\rangle_{\Theta}.
    \end{equation}
    
\begin{figure}
        \centering
        \includegraphics[width=0.75\linewidth]{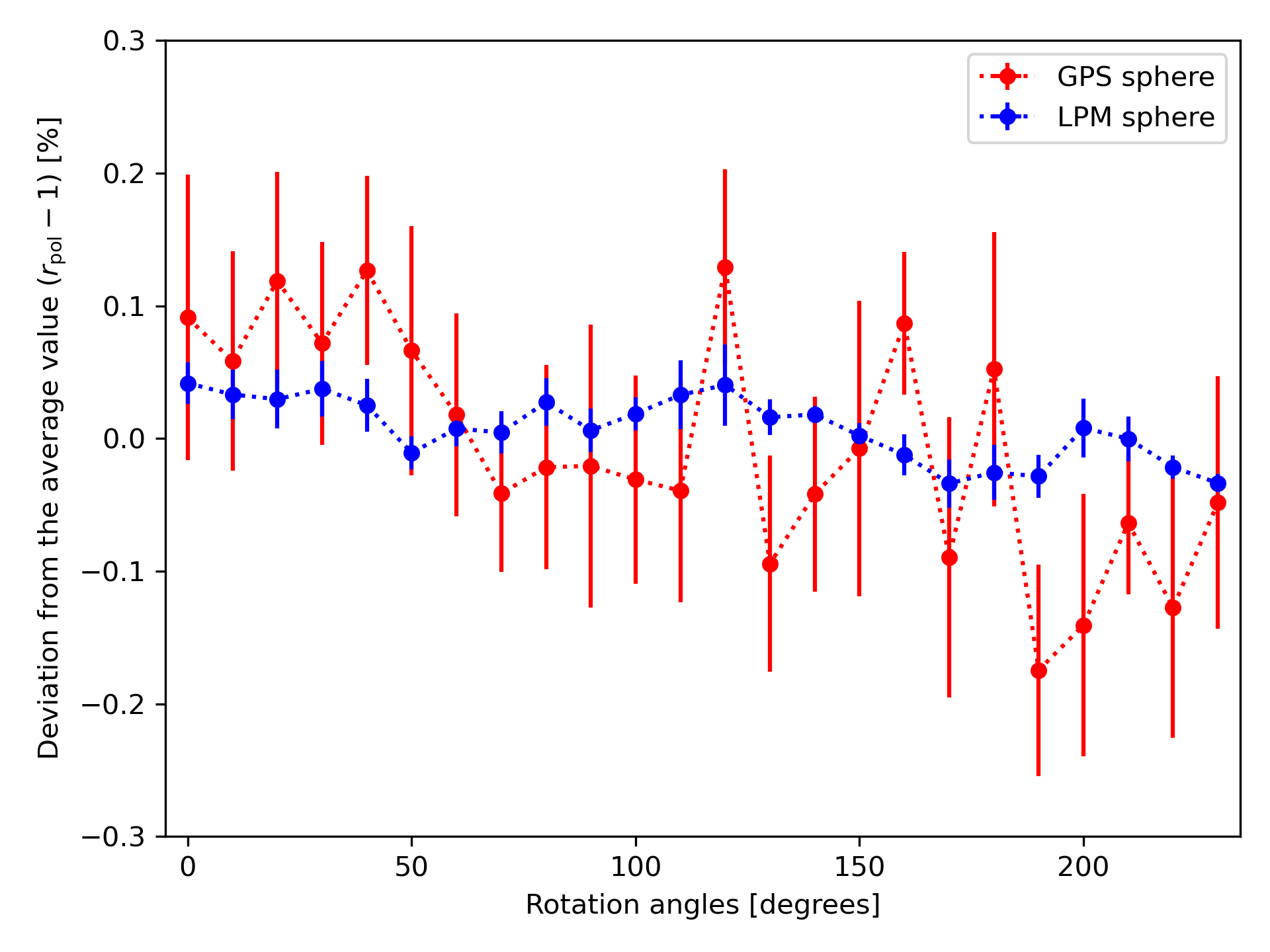}
        \caption{Polarization dependence of the integrating-sphere response for the GPS and LPM spheres. The response was measured from $0^\circ$ to $230^\circ$ in $10^\circ$ increments.}
        \label{fig:polarization}
    \end{figure}

    The response normalized by the mean value over all measured polarization angles is then defined as
    \begin{equation}
        r_{\mathrm{pol}}(\Theta)
        =
        \frac{R_{\mathrm{pol}}(\Theta)}
             {\overline{R}_{\mathrm{pol}}}.
    \end{equation}
    The resulting $r_{\mathrm{pol}}(\Theta)-1$, calculated from the measurement data, is shown in Fig.~\ref{fig:polarization}.

    The relative uncertainty due to the polarization dependence is defined as
    \begin{equation}
        u_{\mathrm{name,pol}}
        =
        s\left(
            r_{\mathrm{pol}}(\Theta)
        \right)_{\Theta},
    \end{equation}
    where ``name'' denotes either the LPM or GPS integrating sphere.
    The resulting uncertainties were
    \begin{align}
        u_{\rm GPS, pol} &= 0.087 \% \label{eq:u_GPS_pol},\\
        u_{\rm LPM, pol} &= 0.024 \% \label{eq:u_LPM_pol}.
    \end{align}
    For both the LPM sphere and the GPS sphere, no clear polarization dependence is evident from these measurements. These standard deviations were therefore used as conservative upper limits on the polarization dependence of the GPS and LPM spheres.

\subsection{Beam-position dependence}
\label{subsec:beam-position_dependence}
    \par If the response of an integrating-sphere power sensor depends on the incident beam position, small variations in the beam position in repeated power-sensor calibration measurements can introduce an additional uncertainty in the measured laser power.
    Therefore, the spatial uniformity of the integrating-sphere response is an important property for evaluating the Pcal laser-power uncertainty.
    
    \par A related spatial non-uniformity measurement of an integrating-sphere transfer standard was reported by Spidell et al. for a NIST--PTB bilateral comparison related to LIGO calibration~\cite{Spidell_NEWRAD2021_NIST_PTB_LIGO}.
    They showed that the spatial non-uniformity of a 100-mm fluoropolymer-lined integrating sphere was much smaller than that of thermopile transfer standards.
    Although the sphere used in that study is expected to be similar to the LPM-type sphere used as the GSK in the KAGRA calibration chain, the detailed implementation of the power sensor, such as the photodetector mounting, port configuration, and readout structure, may not be identical.
    In addition, the GPS-type integrating sphere used in the KAGRA calibration chain was not characterized in that study.
    Therefore, we performed independent measurements of the beam-position dependence for both the LPM- and GPS-type integrating spheres used in the KAGRA calibration chain.

    \par
\begin{figure}
        \centering
        \includegraphics[width=0.90\linewidth]{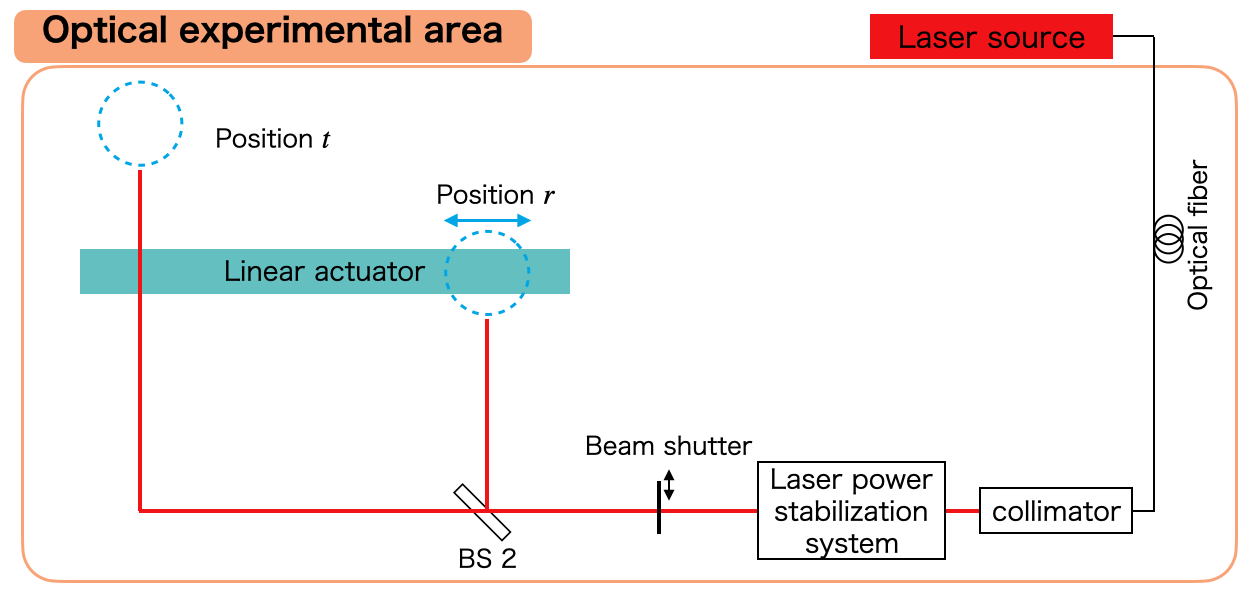}
        \caption{Experimental setup for the beam-position dependence measurement. The transmitted beam is directed to the integrating sphere at position $t$ and used as a reference, while the reflected beam is used to measure the position dependence of the test integrating sphere.}
        \label{fig:position_setup}
    \end{figure}

\begin{figure}
        \centering
        \begin{subfigure}{0.42\linewidth}
            \centering
            \includegraphics[width=\linewidth]{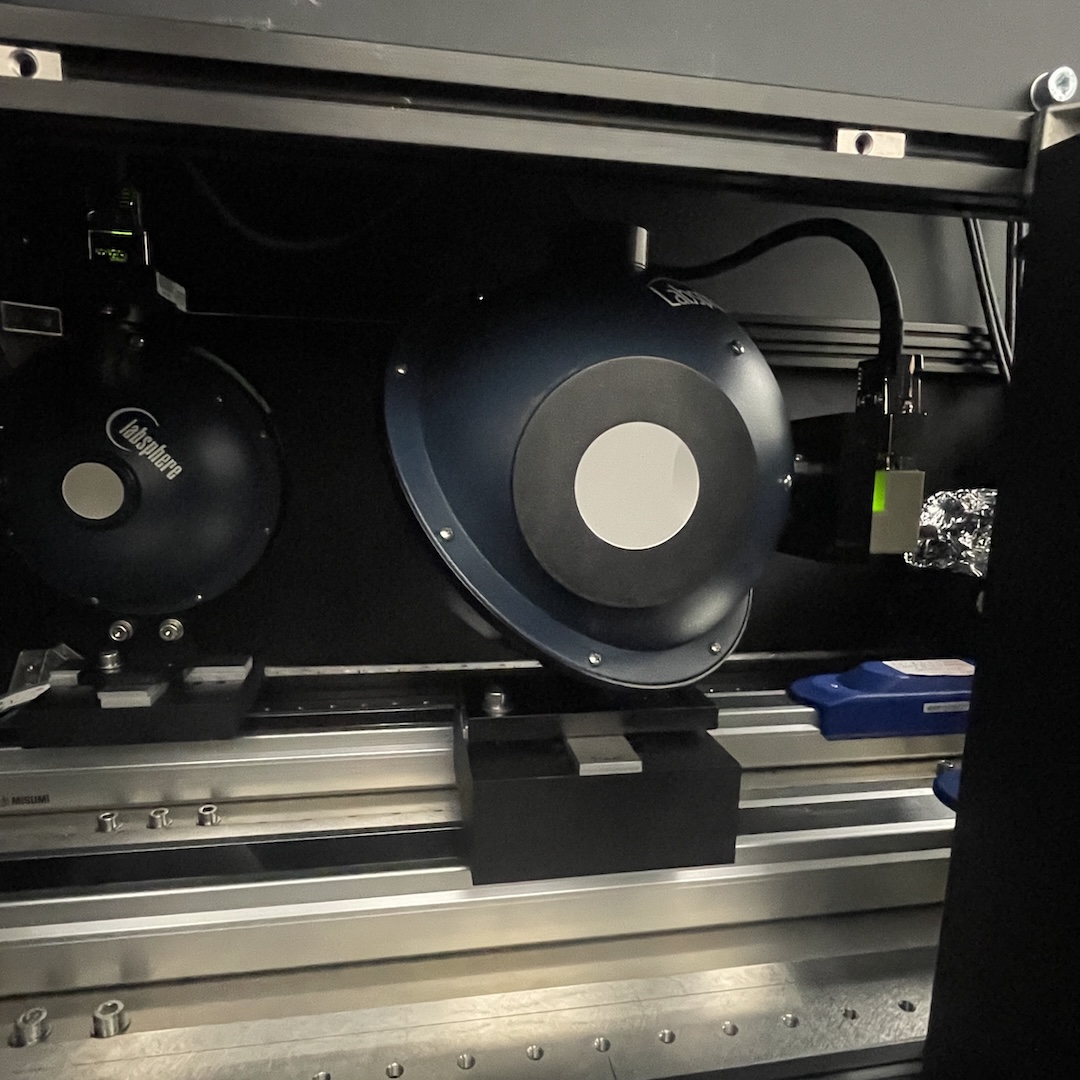}
            \caption{Integrating sphere mounted on the linear actuator.}
        \end{subfigure}
        \hfill
        \begin{subfigure}{0.42\linewidth}
            \centering
            \includegraphics[width=\linewidth]{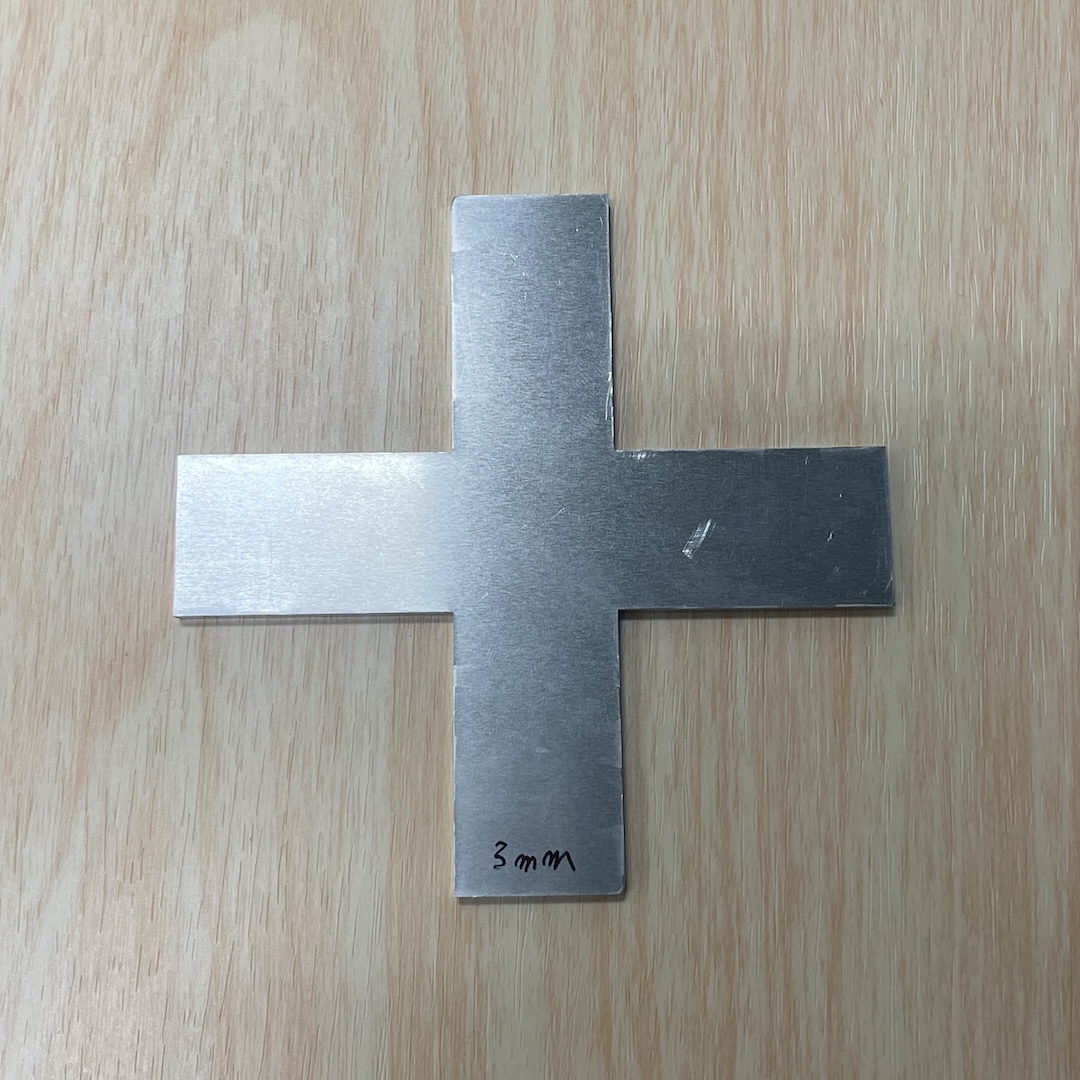}
            \caption{Spacer plates used for vertical adjustment.}
        \end{subfigure}
        \caption{Photographs of the setup used to adjust the incident beam position on the integrating sphere. The sphere is moved horizontally using the linear actuator shown in (a), while its vertical position is adjusted using the spacer plates shown in (b). These adjustments allow the incident beam position to be scanned across the input port in the horizontal and vertical directions.}
        \label{fig:plate}
    \end{figure}

 We used the setup shown in Fig.~\ref{fig:position_setup} to measure the beam-position dependence for the two types of integrating spheres.
    The test integrating sphere was placed at position $r$ on a linear actuator (RS106B-C1-P-3-400-H-S-T-MJ5-KJ3, MISUMI), while the reference integrating sphere was fixed at position $t$ to monitor fluctuations in the input laser power, as in the other measurements.
    The diameter of the laser beam incident on the test integrating sphere was approximately 2.5~mm.
    The test integrating sphere was moved horizontally using the linear actuator, and its vertical position was adjusted in 1~mm increments by placing plates underneath the sphere, as shown in Fig.~\ref{fig:plate}.
    
\begin{figure}
        \centering
        \begin{subfigure}{0.48\linewidth}
            \centering
            \includegraphics[width=\linewidth]{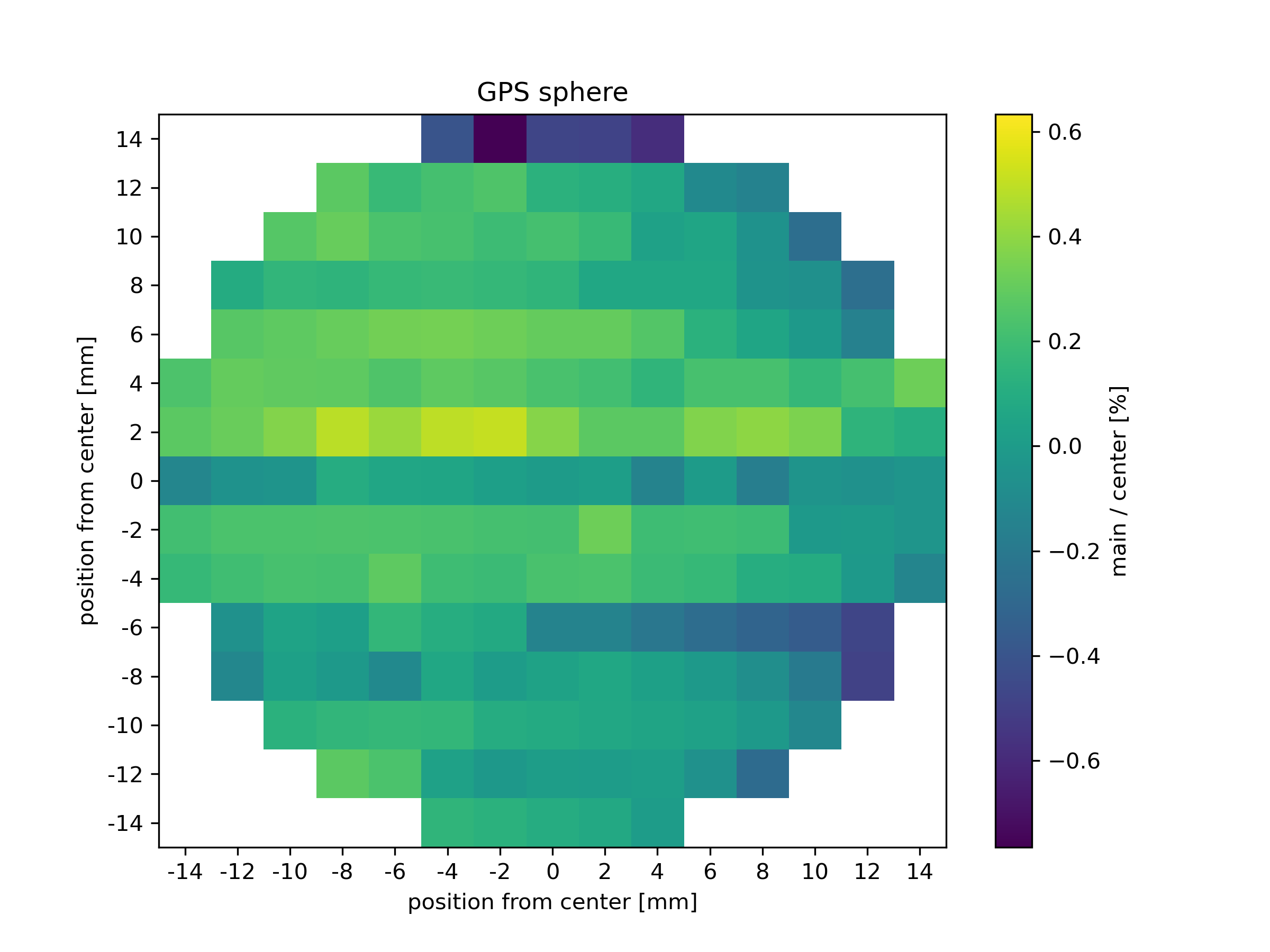}
            \caption{The GPS sphere}
        \end{subfigure}
        \hfill
        \begin{subfigure}{0.48\linewidth}
            \centering
            \includegraphics[width=\linewidth]{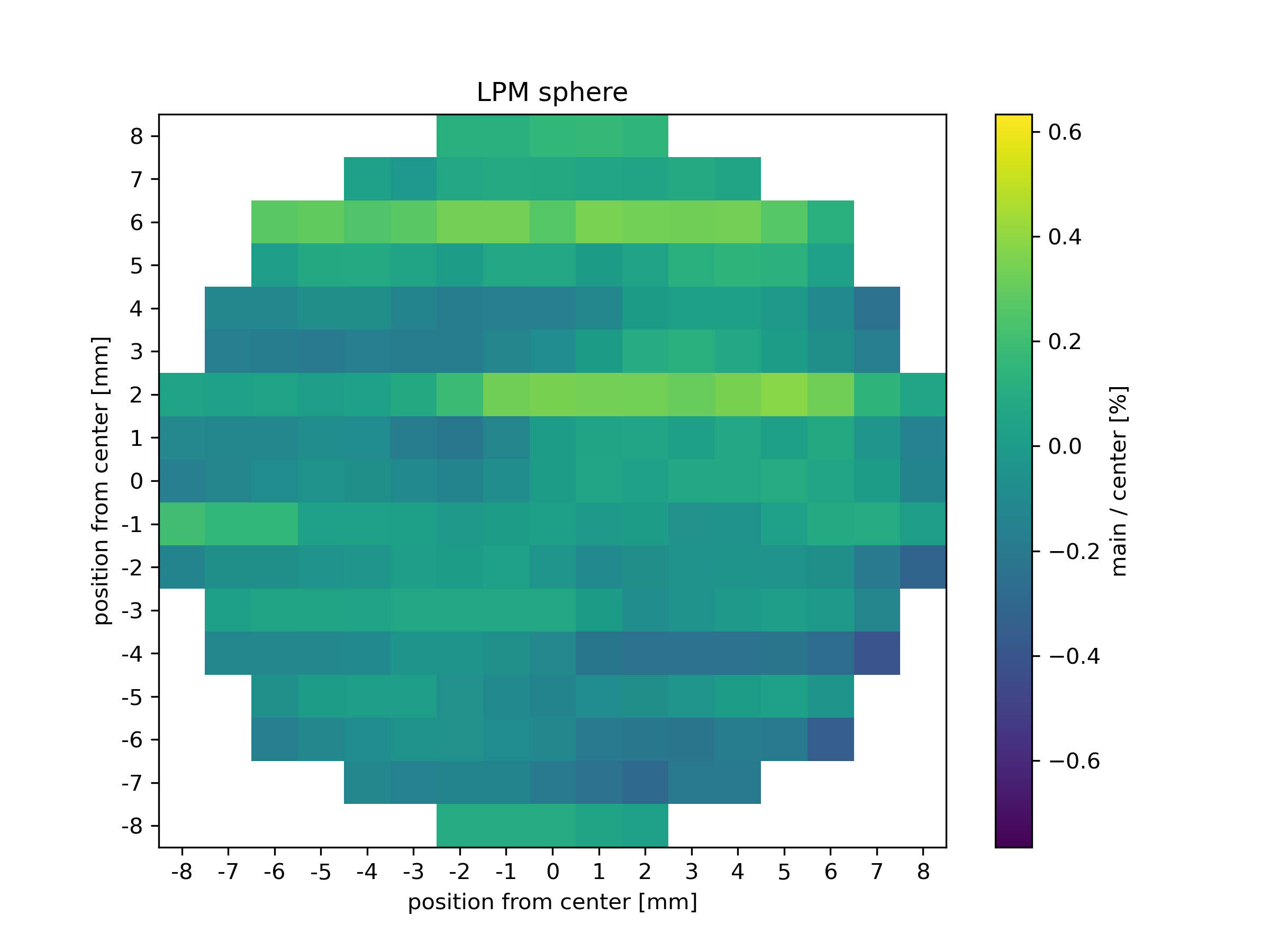}
            \caption{The LPM sphere}
        \end{subfigure}
        \caption{Beam-position dependence of the integrating-sphere response, normalized to the response at the center position. The color scale represents $r_{\mathrm{pos}}-1$ expressed in percent.}
        \label{fig:position}
    \end{figure}

\begin{figure}
        \centering
        \begin{subfigure}{0.48\linewidth}
            \centering
            \includegraphics[width=\linewidth]{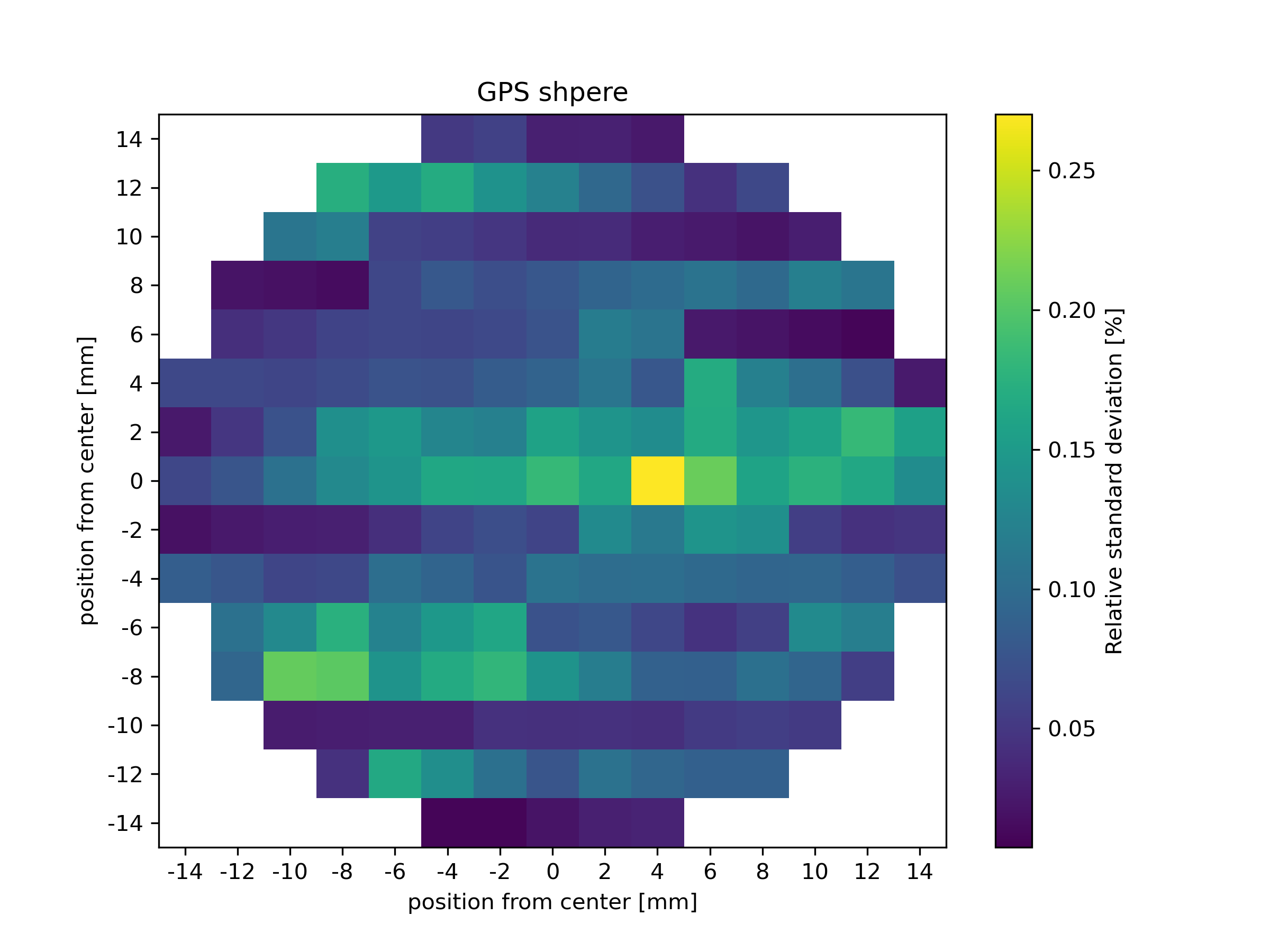}
            \caption{The GPS sphere}
        \end{subfigure}
        \hfill
        \begin{subfigure}{0.48\linewidth}
            \centering
            \includegraphics[width=\linewidth]{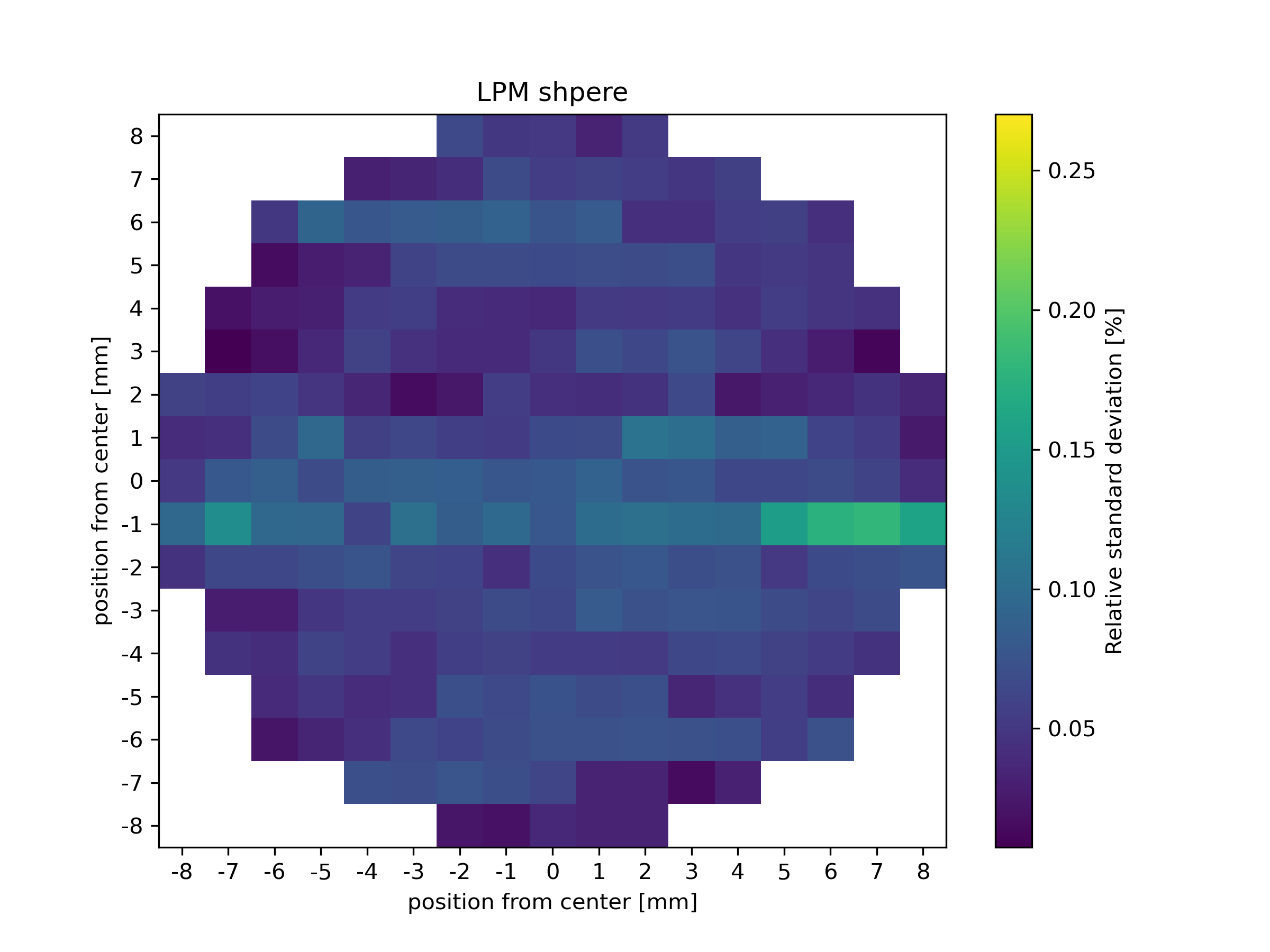}
            \caption{The LPM sphere}
        \end{subfigure}
        \caption{Relative standard deviation of the integrating-sphere response at each measured beam position.}
        \label{fig:position_std}
    \end{figure}

    % Warming up時間に関する記述が欲しい。
    The measurement procedure was as follows.
    \begin{enumerate}
        \item The laser and measurement system were turned on and allowed to stabilize for approximately 1 h.
        \item The test integrating sphere was moved to the center position.
        \item The shutter was closed, and the outputs of the integrating spheres were recorded to obtain the dark-offset data at the center position 
        ($\boldsymbol{V}_{\rm c,d}, \boldsymbol{V}_{\rm ref,c,d}$).
        A total of 30 data points were acquired over approximately 10 s.
        \item The shutter was opened, and the outputs of the integrating spheres were recorded at the center position 
        ($\boldsymbol{V}_{\rm c}, \boldsymbol{V}_{\rm ref,c}$).
        A total of 100 data points were acquired over approximately 30 s.
        \item The vertical position of the test integrating sphere was changed.
        \item The laser was kept incident on the integrating sphere for 10 min to stabilize the temperature of the integrating sphere.
        \item The shutter was closed, and the outputs of the integrating spheres were recorded to obtain the dark-offset data 
        ($\boldsymbol{V}_{\rm pos,d}, \boldsymbol{V}_{\rm ref,pos,d}$).
        A total of 30 data points were acquired over approximately 10 s.
        \item The outputs were recorded while the test integrating sphere was translated back and forth three times, starting from the left, in 1 mm increments for the LPM sphere and 2 mm increments for the GPS sphere 
        ($\boldsymbol{V}_{\rm pos}$ and $\boldsymbol{V}_{\rm ref,pos}$).
        A total of 100 data points were acquired over approximately 30 s.
        \item The shutter was closed again, and the outputs of the integrating spheres were recorded to obtain the final dark-offset data 
        ($\boldsymbol{V}_{\rm pos,d}, \boldsymbol{V}_{\rm ref,pos,d}$).
        A total of 30 data points were acquired over approximately 10 s.
    \end{enumerate}
    The above procedure was repeated at different positions of the test integrating sphere. 
    The position of the test integrating sphere was varied from -8 mm to +8 mm in both the horizontal and vertical directions for the LPM, and from -14 mm to +14 mm for the GPS.
    The response at the center position was calculated as follows:
    \begin{align}
        R_{\mathrm{c}}
        &=
        \left\langle
        \frac{
        \boldsymbol{V}_{\rm c} - \left\langle \boldsymbol{V}_{\rm c, \mathrm{d}} \right\rangle
        }{
        \boldsymbol{V}_{\mathrm{ref},\rm c} - \left\langle \boldsymbol{V}_{\mathrm{ref},\rm c,\mathrm{d}} \right\rangle
        }
        \right\rangle.        
    \end{align}
    For scan lines including the center position, $R_{\mathrm{c}}$ was obtained by averaging the data measured at the center during the corresponding main scan, rather than from a separate center-position measurement.
    The output at each position was normalized to the output at the center position as follows:
    
    \begin{align}
        R_{i, \mathrm{pos}}
        &=
        \left\langle
        \frac{
        \boldsymbol{V}_{i, \rm pos} - \left\langle \boldsymbol{V}_{\rm pos, \mathrm{d}} \right\rangle
        }{
        \boldsymbol{V}_{i,\mathrm{ref},\rm pos} - \left\langle \boldsymbol{V}_{\mathrm{ref},\rm pos,\mathrm{d}} \right\rangle
        }
        \right\rangle, \\
        R_{\mathrm{pos}}
        &=
        \left\langle
        R_{i,\mathrm{pos}}
        \right\rangle_{i=1,\cdots, 6}, \\
        r_{\rm pos}&=
        \frac{
        {R}_{\rm pos}}{
        {R}_{\mathrm{c}}
        },
    \end{align}
    where $i = 1, \cdots, 6$ corresponds to the six measurements obtained from three back and forth scans.
    The relative deviation from the center response, $r_{\mathrm{pos}}-1$, expressed in percent, is shown in Fig.~\ref{fig:position}.
    The standard deviation normalized by the mean at each measurement point is shown in Fig.~\ref{fig:position_std}.
    The LPM output varied within $-0.2\%$ to $+0.2\%$ relative to the center position, and the relative standard deviation at each measurement point was approximately $0.05\%$. In contrast, the GPS output varied within $-0.6\%$ to $+0.6\%$, with a relative standard deviation ranging from $0.05\%$ to $0.2\%$.
    The variation in the output exceeds the measurement uncertainty, suggesting that the observed position dependence is not solely due to measurement uncertainty.
    The observed sub-percent position dependence of the LPM-type sphere is consistent with the spatial non-uniformity reported for the integrating-sphere transfer standard in Ref.~\cite{Spidell_NEWRAD2021_NIST_PTB_LIGO}.
    \par Based on these measurements, the uncertainty associated with the beam position was estimated.
    The relative uncertainty due to the beam-position dependence for a spatial
    evaluation range $\mathcal{D}$ is defined as
    \begin{equation}
        u_{\mathrm{name,pos}|\mathcal{D}}
        =
        s\left(
            r_{\mathrm{pos}}
        \right)_{\mathrm{pos}\in\mathcal{D}},
    \end{equation}
    where ``name'' denotes either the LPM or GPS integrating sphere.
    Using all measured positions within the scanned area, the relative uncertainties were evaluated as
    \begin{align}
    u_{\rm GPS,pos|full} &= 0.21\%,\\
    u_{\rm LPM,pos|full} &= 0.15\%.
    \end{align}
    These values characterize the overall position dependence of each type of integrating sphere over the full measured area.

    \par In practical calibration measurements, the relevant position-dependent uncertainty depends on the spatial range over which the incident beam can vary. The corresponding range was therefore evaluated for each calibration configuration when applying the measured position dependence to the O4a Pcal-X data, as described in Sec.~\ref{sec:application}.

\section{Treatment of integrating-sphere systematic uncertainties in Pcal system}
\label{sec:Uncertainty_modeling}
    \par
    Based on the measurements described in Sec.~\ref{sec:Experimental} and the calibration procedures described in Subsection~\ref{subsec:Pcal_power_estimation},
    we assigned systematic uncertainty contributions associated with incidence angle, polarization, and beam position to each calibration configuration.
    The applicable contributions depend on whether the integrating sphere or power sensor was repositioned during calibration and on whether the incident polarization was controlled.
    No additional systematic uncertainty from these effects was assigned to the absolute GSK calibration at LIGO, which serves as the reference for the subsequent calibration chain.
    For the GSK–WSK comparison at U. Toyama, incidence-angle and beam-position uncertainties were included because both integrating spheres were manually repositioned, whereas no polarization uncertainty was included because the incident polarization was controlled with a PBS to maintain p polarization, consistent with the GSK calibration at LIGO. At the KAGRA end station, all three contributions were included for measurements using WSK because its position and orientation were manually adjusted and the incident polarization was not well controlled owing to the complex optical paths in the Tx module and the mirror configuration in the EA chamber.
    For the RxPD, incidence-angle and beam-position uncertainties were included because it was repositioned during calibration,
    whereas no polarization uncertainty was included because its orientation was unchanged.
    No additional systematic uncertainty was assigned to the TxPDs because they remained fixed during the calibration procedure.
    These assignments are summarized in Table~\ref{tab:uncertainty_summary}.

    \begin{table}[htbp]
        \centering
        \caption{Summary of systematic uncertainty contributions from angle, position, and polarization at each measurement location and power sensor.}
        \label{tab:uncertainty_summary}

        \small
        \setlength{\tabcolsep}{4pt}
        \renewcommand{\arraystretch}{1.15}

        \begin{tabular}{lccccccc}
            \hline
            Place
            & \multicolumn{1}{c}{LIGO}
            & \multicolumn{2}{c}{U. Toyama}
            & \multicolumn{4}{c}{End station} \\
            \hline

            Power sensor
            & GSK
            & GSK
            & WSK
            & \makecell{WSK\\at Tx}
            & \makecell{WSK\\at Rx}
            & RxPD
            & TxPD \\
            \hline

            Incidence-angle uncertainty
            & No & Yes & Yes & Yes & Yes & Yes & No \\

            Polarization uncertainty
            & No & No & No & Yes & Yes & No & No \\

            Position uncertainty
            & No & Yes & Yes & Yes & Yes & Yes & No \\
            \hline
        \end{tabular}
    \end{table}

    \par We incorporated these systematic uncertainties into the Pcal calculations described in Subsection~\ref{subsec:Pcal_power_estimation}.
    Here, the systematic uncertainties $u_{\rm name,ang}$, $u_{\rm name,pol}$, and $u_{\rm name,pos}$, as well as their combined quantities $u_{s1}$--$u_{s4}$, are relative uncertainties, whereas $u(X)$ and $u_0(X)$ denote absolute uncertainties of the quantity $X$.
    First, for the GSK--WSK comparison at U. Toyama, the relevant systematic uncertainty contributions were combined in quadrature with the original uncertainty of each responsivity-ratio measurement $\alpha_{i,{\rm WSK/GSK}}$ defined in Eq.~\ref{eq:alpha_i}.
    Denoting the original uncertainty by $u_0(\alpha_{i,{\rm WSK/GSK}})$, the updated uncertainty is given by
    \begin{equation}
        u(\alpha_{i, {\rm WSK/GSK}}) = \sqrt{(u_0(\alpha_{i, {\rm WSK/GSK}}))^2 + ( \alpha_{i, {\rm WSK/GSK}}u_{s1} ) ^2},
    \end{equation}
    where,
    \begin{equation}
        u^2_{s1} = u^2_{\rm LPM, ang|0} + u^2_{\rm LPM, pos, GSK} + u^2_{\rm GPS, ang|0} + u^2_{\rm GPS, pos, WSK@U.Toyama}
    \end{equation}
    For $\alpha_{\mathrm{RxPD/WSK},j}$ and $\alpha_{\mathrm{TxPD}j/\mathrm{WSK}}$, the corresponding systematic uncertainty contributions are added to the original uncertainties as follows.

    \begin{align}
        u(\alpha_{\mathrm{RxPD/WSK},j}) &= \sqrt{(u_0(\alpha_{\mathrm{RxPD/WSK},j}))^2 + (\alpha_{\mathrm{RxPD/WSK},j}u_{s2,j})^2}, \\
        u_{s2,j}^2 &= u^2_{{\rm GPS, ang}|\theta_j} + u^2_{\rm GPS, pos, RxPD} + u^2_{\rm GPS, ang|0} + u^2_{\rm GPS, pol} + u^2_{\rm GPS, pos, WSK@Rx}, \\
        u(\alpha_{\mathrm{TxPDj/WSK}}) &= \sqrt{(u_0(\alpha_{\mathrm{TxPDj/WSK}}))^2 + (\alpha_{\mathrm{TxPDj/WSK}} u_{s3})^2}, \\
        u_{s3}^2 &= u^2_{\rm GPS, ang|0} + u^2_{\rm GPS, pol} + u^2_{\rm GPS, pos, WSK@Tx}.
    \end{align}
    Because the incidence angles of the two laser paths on the RxPD differ, $u^2_{s2,j}$ depends on the incidence angle $\theta_j$ of each path.
    Similarly, because the optical efficiencies $e_j$ are determined from measurements using WSK at the Rx and Tx locations, their uncertainties are also affected by the corresponding systematic uncertainty contributions:
    \begin{align}
        u(e_j) &= \sqrt{(u_0(e_j))^2 + (e_j u_{s4})^2}, \\
        u_{s4}^2 &= 2u^2_{\rm GPS, ang|0} + 2u^2_{\rm GPS, pol} + u^2_{\rm GPS, pos, WSK@Rx} + u^2_{\rm GPS, pos, WSK@Tx}.
    \end{align}
    The position-dependent uncertainty terms in these equations depend on the beam-position variation associated with each calibration configuration.
    Their values for the O4a Pcal-X measurements are evaluated in Sec.~\ref{sec:application}.

\section{Application of the uncertainty model to O4a Pcal-X data}
\label{sec:application}
    \par We applied the systematic uncertainties evaluated above to the calibration measurements of KAGRA Pcal-X during O4a~\cite{Chen_2025}.
    The polarization-dependent uncertainties used in this analysis are those evaluated in Sec.~3.2 and given in Eqs.~\ref{eq:u_GPS_pol} and \ref{eq:u_LPM_pol}.
    For the incidence-angle dependence, the uncertainty was evaluated separately for each calibration configuration according to its nominal incidence angle.
    For measurements in which the integrating sphere was aligned for normal incidence, the nominal angle was taken to be $0^\circ$.
    In contrast, the two Pcal beams incident on the RxPD entered at different angles.
    Measurements performed at the KAGRA end station showed that the incidence angles on the RxPD were approximately $5^\circ$ for path 1 and $-4^\circ$ for path 2, and these measured values were therefore used as the nominal incidence angles in the uncertainty evaluation.
    For each nominal angle, an alignment reproducibility of $\pm2^\circ$ was assumed, and the corresponding incidence-angle uncertainty was evaluated using the method described in Sec.~3.1.
    The resulting relative uncertainties due to the incidence-angle variation are summarized in Table~\ref{tab:incident_angle_result}.

    \begin{table}[htbp]
    \centering
    \caption{Relative uncertainties due to incidence-angle variation used in the O4a Pcal-X calibration}
    \begin{tabular}{lccc}
        \hline
        Incidence angle $\theta_0$ [degree] & -4 & 0 & 5 \\
        \hline
        $u_{\rm GPS, ang|\theta_0}$ & 0.050\% & 0.087\% & 0.073\%  \\
        $u_{\rm LPM, ang|\theta_0}$ & - & 0.020\% & -  \\
        \hline
    \end{tabular}
    \label{tab:incident_angle_result}
    \end{table}

    \par Section~\ref{subsec:beam-position_dependence} characterized the position dependence over the full measured area. For the application to the O4a calibration data, the uncertainty was instead evaluated using the range of beam-position variations expected for each calibration configuration.

    \par For all measurements in which an integrating sphere was manually positioned, a $\pm 2$ mm positional variation was assumed. In addition, a $\pm 2^\circ$ angular misalignment was assumed, and the corresponding beam-center displacement was estimated as $d_{\rm ang}=R\theta$, where $R$ is the radius of the integrating sphere and $\theta$ is the angular misalignment. The direct positional variation is denoted by $d_{\rm pos}$.
    Thus, both the direct positional variation and the displacement induced by angular misalignment were treated as contributions to the variation of the beam-center position.

    \par For measurements performed in the Rx module, an additional beam-center variation observed during O4a was included as $d_{\rm obs}= \pm 8$ mm; otherwise, $d_{\rm obs}$ was set to zero. The beam radius is denoted by $r_{\rm beam}$.

    \par The spatial range used to evaluate the position-dependent uncertainty was then defined as
    \begin{equation}
        d_{\rm eval}
        =
        \sqrt{
        d_{\rm ang}^2
        +
        d_{\rm pos}^2
        +
        d_{\rm obs}^2
        }
        +
        r_{\rm beam}.
    \end{equation}
    The beam-center variations were combined in quadrature, while the beam radius was added linearly because it represents the finite spatial extent illuminated by the beam rather than an independent variation of the beam-center position. The resulting position ranges assumed for each calibration configuration are summarized in Table~\ref{tab:position_range_o4a}.
    These ranges are used to evaluate the corresponding position-dependent uncertainty terms in Sec.~\ref{sec:Uncertainty_modeling}, such as $u_{\rm GPS,pos,WSK@Tx}$ and $u_{\rm GPS,pos,WSK@Rx}$.

    \begin{table}[htbp]
        \centering
        \caption{Position ranges used to evaluate the configuration-specific position-dependent uncertainties for the O4a Pcal-X calibration.}
        \label{tab:position_range_o4a}

        \small
        \setlength{\tabcolsep}{4pt}
        \renewcommand{\arraystretch}{1.15}

        \begin{tabular}{lcccccc}
            \hline
            Power sensor
            & GSK
            & \makecell{WSK at\\U. Toyama}
            & \makecell{WSK\\at Tx}
            & \makecell{WSK\\at Rx}
            & RxPD
            & TxPD \\
            \hline
            Position range
            & $\pm 4$ mm
            & $\pm 4$ mm
            & $\pm 10$ mm
            & $\pm 14$ mm
            & $\pm 14$ mm
            & -- \\
            \hline
        \end{tabular}
    \end{table}
    
    \par For the GSK, which uses the LPM-type integrating sphere, and for the WSK and RxPD, which use the GPS-type integrating sphere, the relative uncertainties obtained from the position-dependence measurements were evaluated as
    \begin{align}
        u_{{\rm LPM,pos},{\rm GSK}} &= u_{\rm LPM,pos|\pm4\,mm} = 0.13\%,\\
        u_{{\rm GPS,pos},{\rm WSK@U.Toyama}} &= u_{\rm GPS,pos|\pm4\,mm} = 0.16\%,\\
        u_{{\rm GPS,pos},{\rm WSK@Tx}} &= u_{\rm GPS,pos|\pm10\,mm} = 0.17\%,\\
        u_{{\rm GPS,pos},{\rm WSK@Rx}} &= u_{{\rm GPS,pos},{\rm RxPD}} = u_{\rm GPS,pos|\pm14\,mm} = 0.21\%.
    \end{align}
    These values were applied to the corresponding calibration configurations according to the position ranges listed in Table~\ref{tab:position_range_o4a}.

    \par
\begin{figure}[htbp]
            \centering
            
            \begin{subfigure}{0.48\linewidth}
                \centering
                \includegraphics[width=\linewidth]{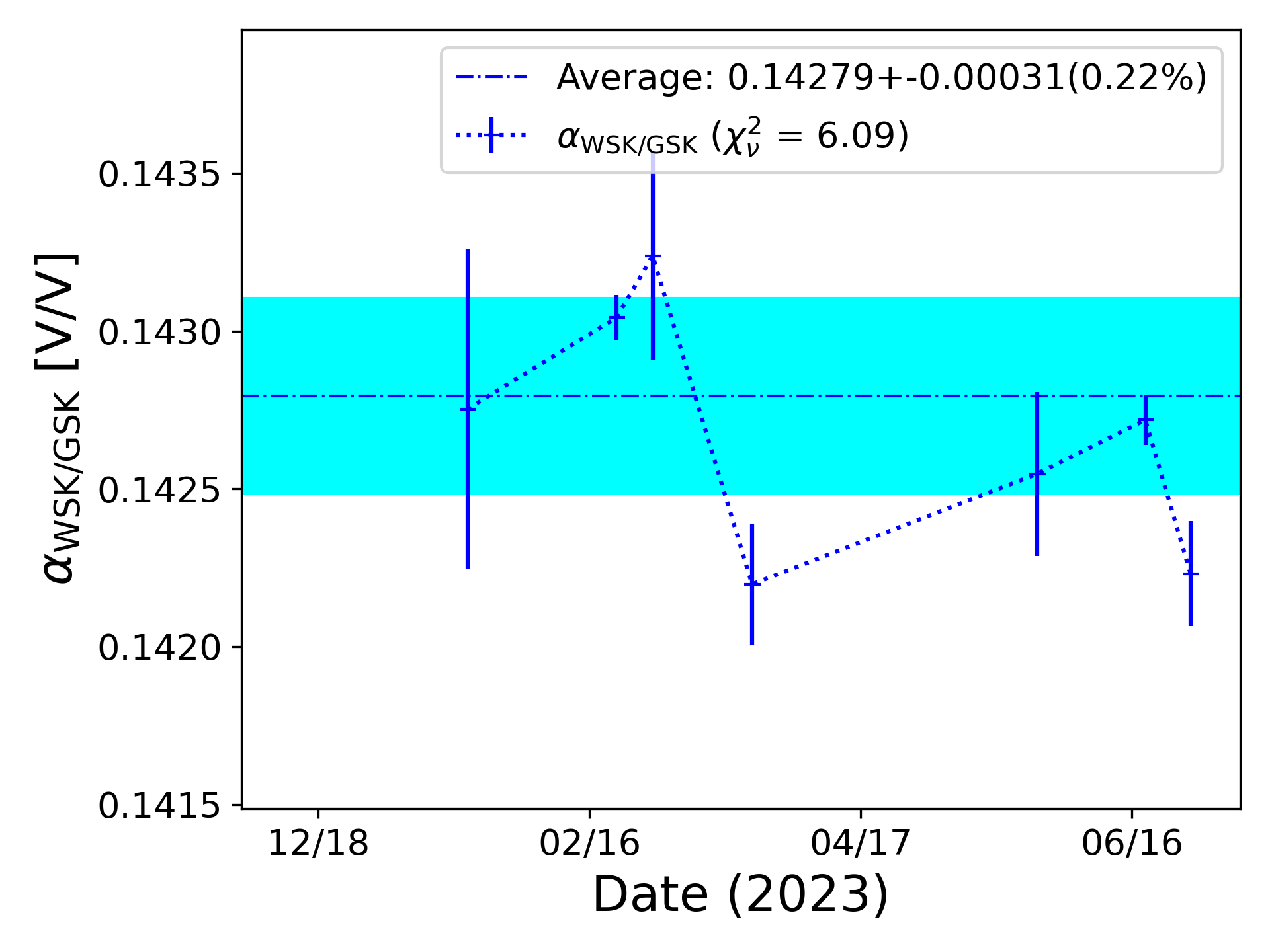}
                %\caption{Caption 1}
            \end{subfigure}
            \hfill
            \begin{subfigure}{0.48\linewidth}
                \centering
                \includegraphics[width=\linewidth]{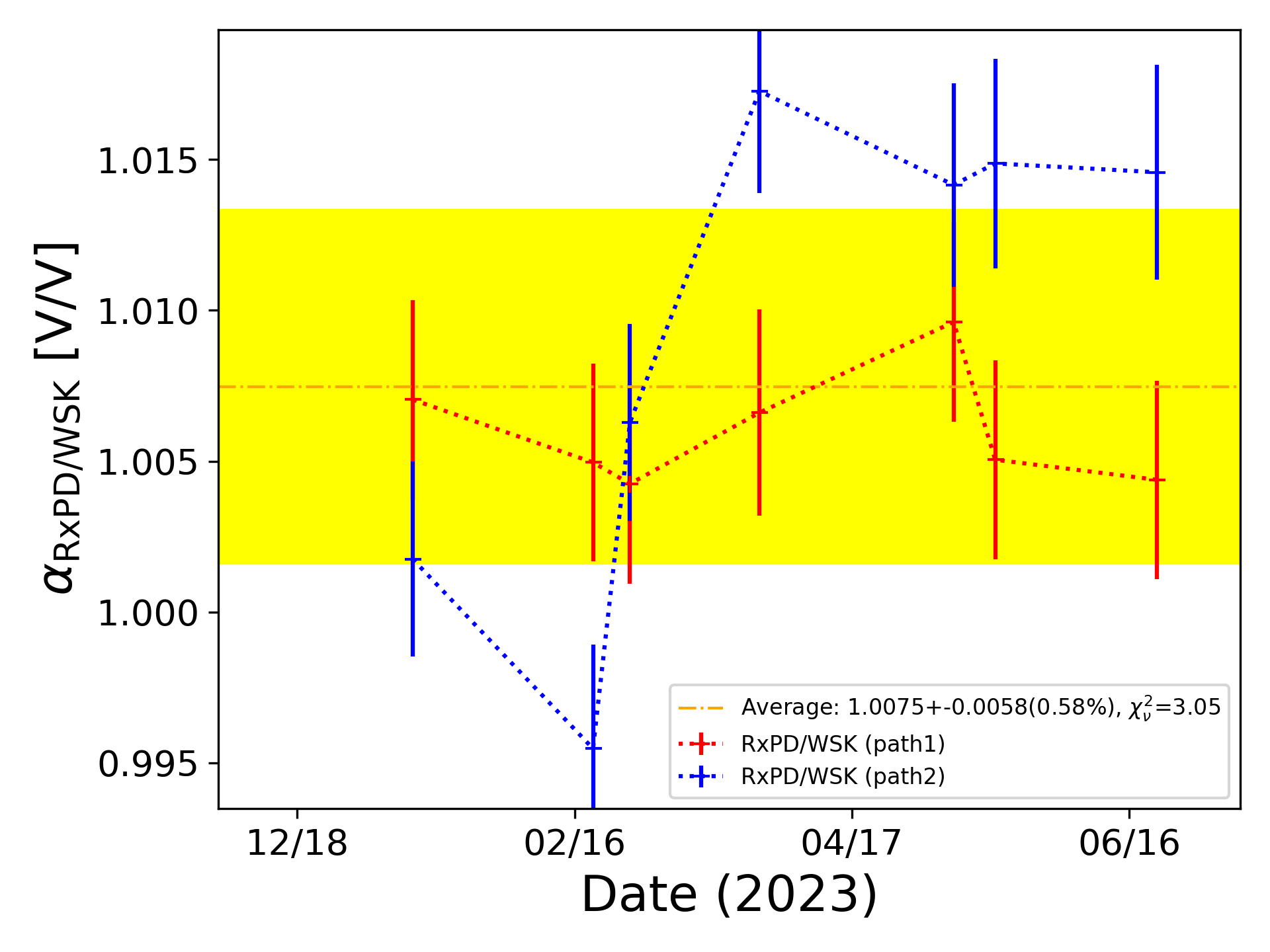}
                %\caption{Caption 2}
            \end{subfigure}
            
            \vspace{0.3cm}
            
            \begin{subfigure}{0.48\linewidth}
                \centering
                \includegraphics[width=\linewidth]{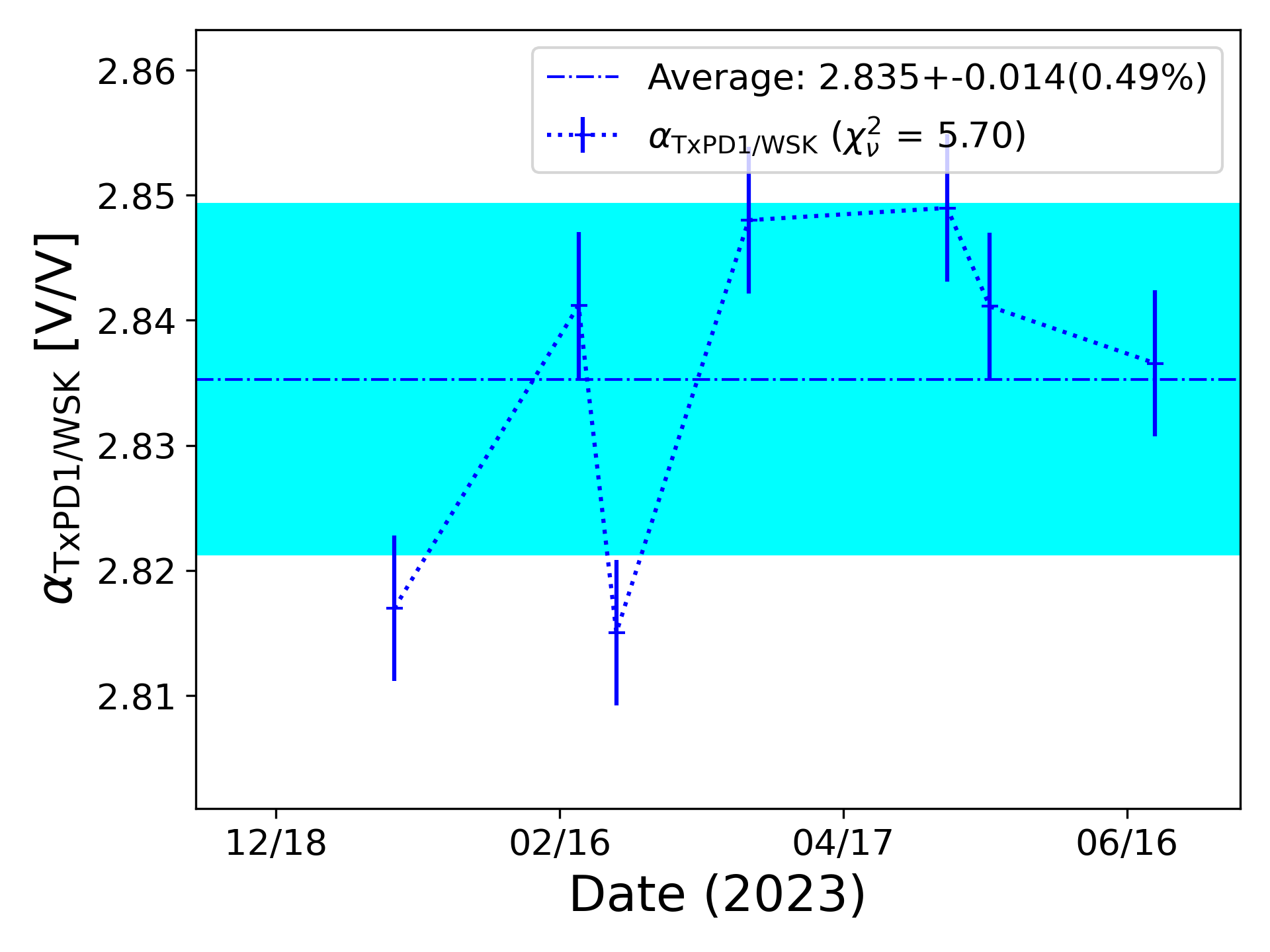}
                %\caption{Caption 3}
            \end{subfigure}
            \hfill
            \begin{subfigure}{0.48\linewidth}
                \centering
                \includegraphics[width=\linewidth]{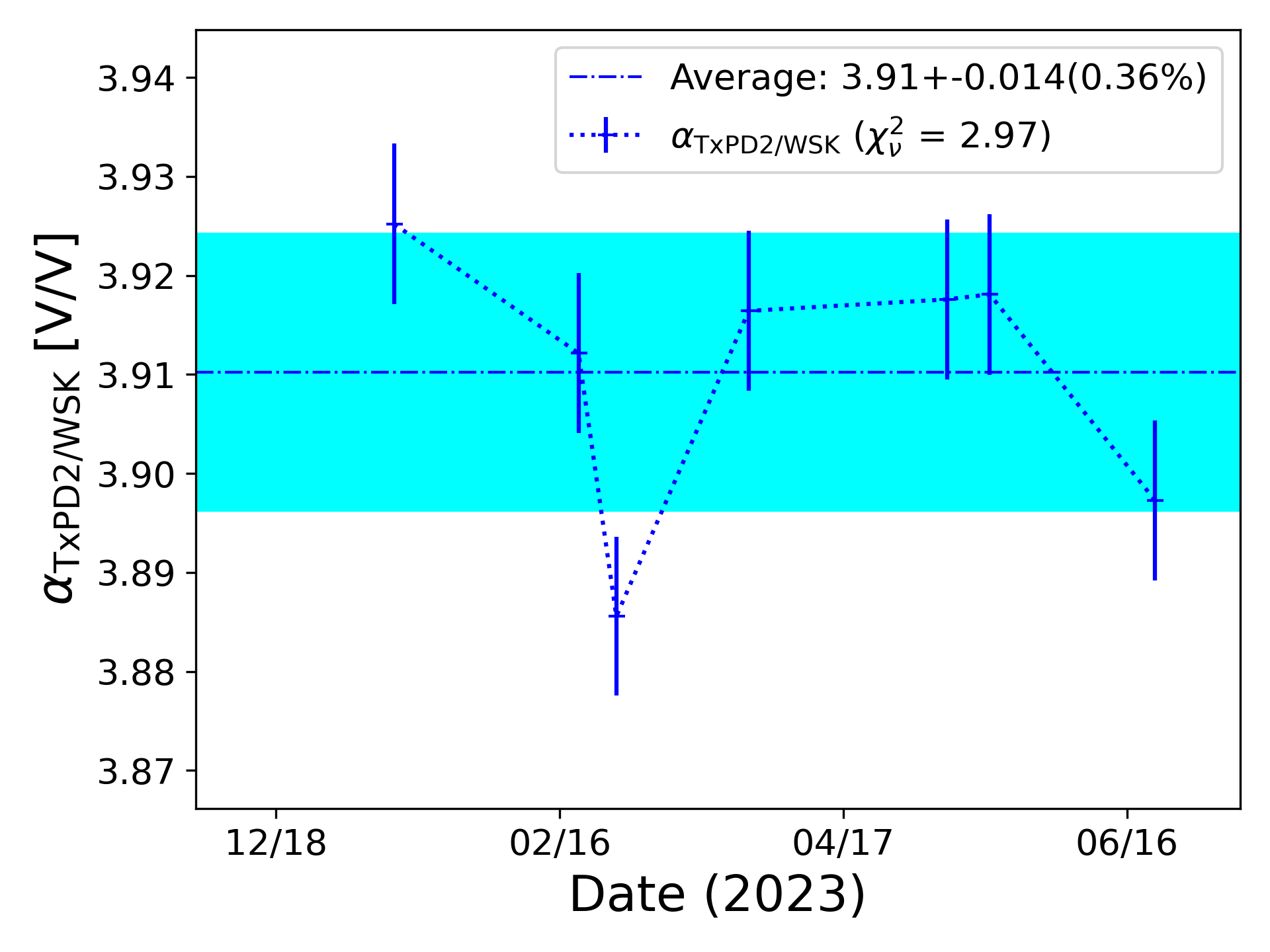}
                %\caption{Caption 4}
            \end{subfigure}

            \vspace{0.3cm}
            
            \begin{subfigure}{0.48\linewidth}
                \centering
                \includegraphics[width=\linewidth]{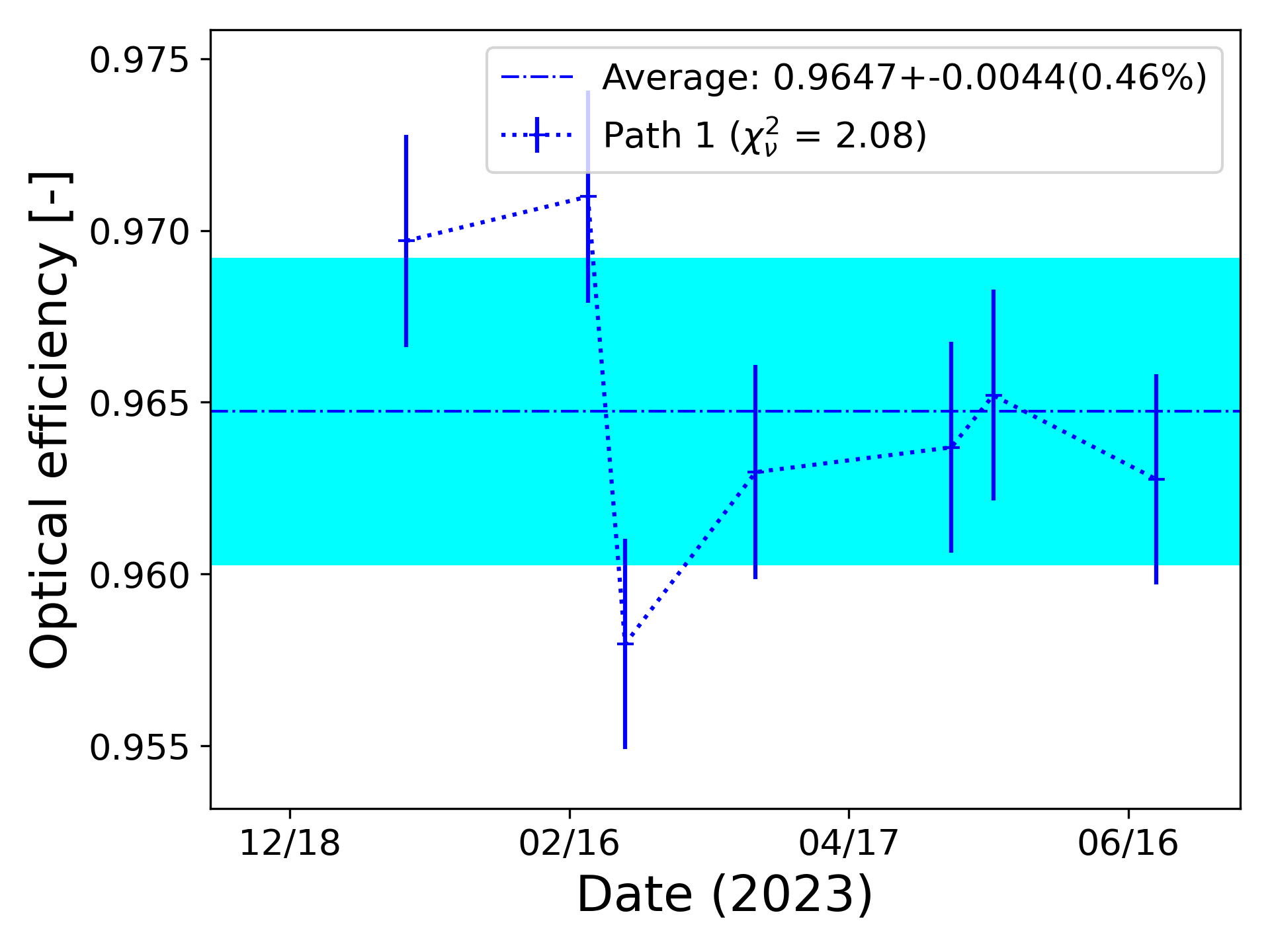}
                %\caption{Caption 5}
            \end{subfigure}
            \hfill
            \begin{subfigure}{0.48\linewidth}
                \centering
                \includegraphics[width=\linewidth]{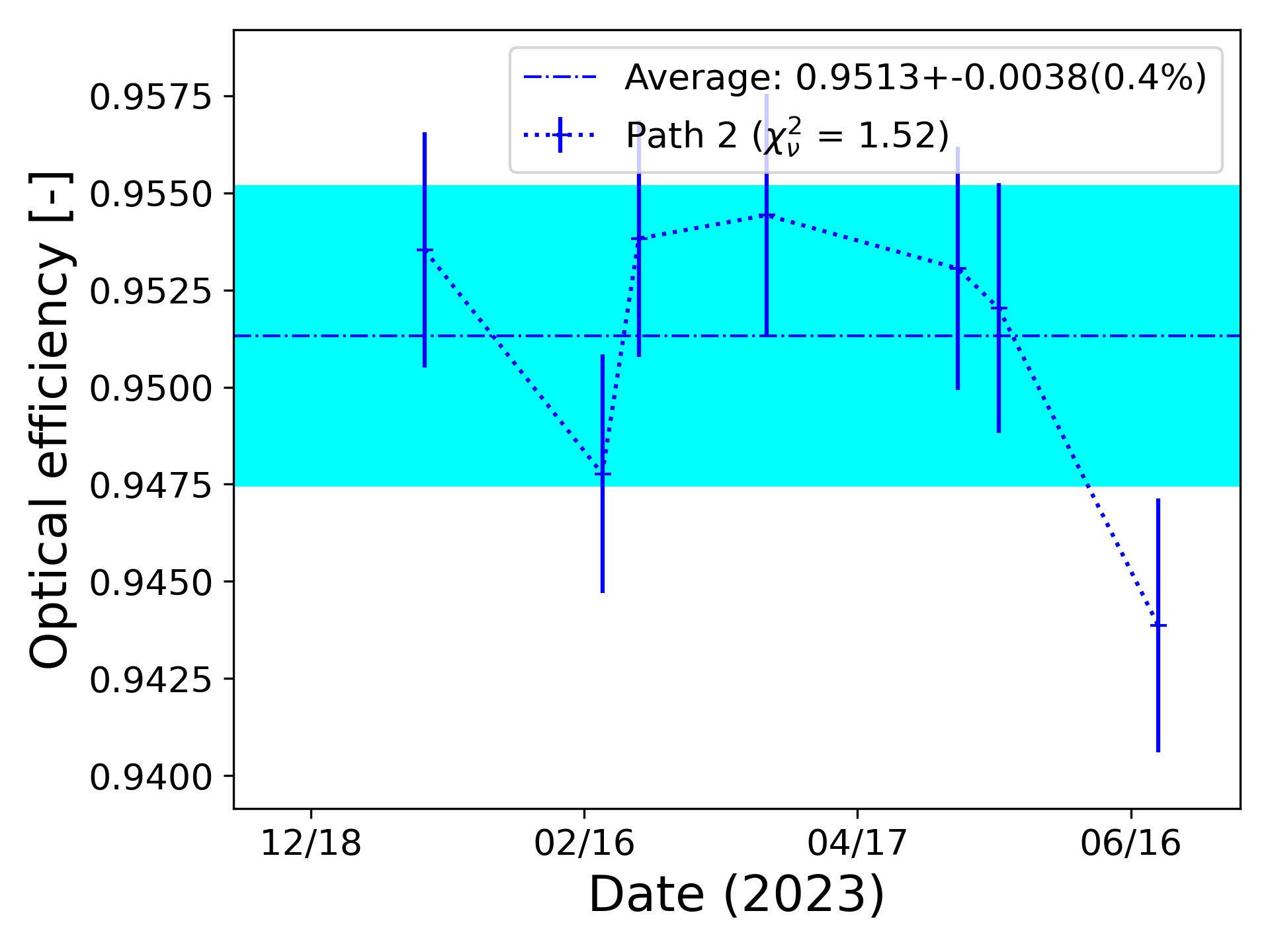}
                %\caption{Caption 6}
            \end{subfigure}
            
            \caption{
                Reanalysis of the O4a calibration measurements previously reported in Ref.~\cite{Chen_2025}.
                The underlying calibration data are the same as those shown in Fig.~\ref{fig:NOSYS}, while the uncertainties of the individual measurements are updated here to include the integrating-sphere systematic uncertainties considered in this work.
                Because the weighted averages depend on the individual uncertainties, updating the uncertainties can also slightly shift the averaged values.
            }
            \label{fig:with_systematic_uncertainty}
        \end{figure}

 The updated responsivity ratios are shown in Fig.~\ref{fig:with_systematic_uncertainty}.
    Compared to Fig.~\ref{fig:NOSYS}, the error bars for $\alpha_{\rm WSK/GSK}$ data, measured at the U. Toyama, remained largely unchanged, whereas the error bars for $\alpha_{\rm RxPD/WSK}, \alpha_{\rm TxPD1/WSK}$, and $\alpha_{\rm TxPD2/WSK}$, measured at the end station, became significantly larger.

    \par Calculations were performed with and without the integrating-sphere systematic uncertainties, and the resulting reduced chi-square values $\chi_\nu^2$ are summarized in Table~\ref{tab:reduced_chi_squre}.

    \begin{table}[htbp]
    \centering
    \caption{Reduced chi-square values $\chi_\nu^2$ with and without integrating-sphere systematic uncertainties}
    \begin{tabular}{lcccccc}
        \hline
         & $\alpha_{\rm WSK/GSK}$ & $\alpha_{\rm RxPD/WSK}$ & $\alpha_{\rm TxPD1/WSK}$ & $\alpha_{\rm TxPD2/WSK}$ & $e_1$ & $e_2$ \\
        \hline
        Without & 6.76 & 59.53 & 3044 & 1199 & 333 & 22.28 \\
        With & 6.09 & 3.05 & 5.70 & 2.97 & 2.08 & 1.52 \\
        \hline
    \end{tabular}
    \label{tab:reduced_chi_squre}
    \end{table}
    
    \par The reduced chi-square values decrease substantially after including the systematic uncertainties, particularly for the parameters measured at the KAGRA end station.
    However, several values remain greater than unity, indicating that the observed scatter is not fully described by the systematic effects considered in this study.

    \par Using the updated uncertainties of the individual calibration measurements, the Pcal-X laser power at the ETMX during O4a was recalculated following the same procedure as in Ref.~\cite{Chen_2025}. The resulting laser power was
    \(P = 2.006 \pm 0.014~\mathrm{W}.\)
    Using this updated power uncertainty together with the other parameters in the Pcal displacement model, the displacement induced by Pcal-X at 1~Hz was evaluated as
    \(x_{1\, \mathrm{Hz}} = (1.480 \pm 0.010)\times10^{-11}~\mathrm{m}.\)
    The corresponding relative uncertainty is 0.70 \%. Compared with the previously reported value of 0.63 \%~\cite{Chen_2025}, the updated uncertainty is slightly larger as a result of explicitly including the systematic uncertainties associated with the incident conditions of the integrating spheres.

\section{Discussion}
\label{sec:discussion}
    \par Although the integrating-sphere systematic uncertainties account for a substantial part of the measurement-to-measurement variation, several reduced chi-square values remain greater than unity. This indicates that additional sources of variation are not captured by the incidence-angle, polarization, and beam-position dependencies considered in the present uncertainty model.
    In particular, the measurements performed at U. Toyama show only a limited improvement, suggesting that additional sources of variation may be present in the calibration setup.

    \par One possible additional source is the temperature dependence of the integrating-sphere responsivity. Changes in the temperature of the photodetector or integrating-sphere assembly could lead to small variations in responsivity between calibration measurements performed on different days. %cite LIGO, Virgoはtemperatureの補正をしているため、その影響があることは引用して補強しても良いかも
    Such effects were not explicitly included in the present uncertainty model and may contribute to the remaining excess scatter.

    \par Another limitation arises from the assumed reproducibility of the integrating-sphere placement. In this study, positional and angular variations of $\pm2$ mm and $\pm2^\circ$, respectively, were assumed for manually positioned integrating spheres. These values represent reasonable estimates of the setup reproducibility, but the actual distributions of the position and angle variations were not measured directly.
    In addition, the beam position and incidence angle may not be completely independent in actual calibration measurements. For example, a small tilt or repositioning of an integrating sphere can change both the incidence angle and the beam position on the input port. In the present uncertainty model, these effects were evaluated separately, and possible correlations between them were not explicitly included.
    If the true variations were larger, had nonuniform distributions, or were correlated with each other, the corresponding systematic uncertainties could be affected.
    
    \par In addition, other instabilities in the calibration apparatus that were not explicitly modeled may also contribute to the remaining scatter. Further measurements that monitor the integrating-sphere temperature and directly record the beam position and incidence angle during each calibration would help distinguish these effects and allow a more accurate evaluation of the remaining uncertainty.

\section{Conclusion}
\label{sec:conclusion}
\par In this study, we investigated systematic effects associated with the incident-beam conditions on integrating-sphere power sensors used in the KAGRA Pcal system. We experimentally characterized the dependence of the integrating-sphere response on the incidence angle, polarization, and beam position for the LPM- and GPS-type integrating spheres used in the KAGRA calibration chain, and incorporated the corresponding systematic uncertainties into the uncertainty model for the O4a Pcal-X calibration measurements.

\par The extended uncertainty model substantially reduced the reduced chi-square values of the repeatedly measured calibration parameters, particularly for measurements performed at the KAGRA end station, indicating that variations in the incident conditions of the integrating spheres account for a substantial part of the previously unexplained excess scatter. Several reduced chi-square values nevertheless remain greater than unity, motivating further characterization of the remaining sources of variation to improve the reliability of Pcal laser-power calibration.

\appendix

\section{Average and uncertainty calculation}
\label{app:average}

\par
In this study, repeated measurements are combined using the same averaging
procedure as that adopted in Ref.~\cite{Chen_2025}.
When each measured value $A_i$ ($i=1,\ldots,N$) has an associated
uncertainty $\sigma_i$, the weighted average is calculated using the
inverse variance as the weight,
\begin{equation}
    w_i = \frac{1}{\sigma_i^2},
\end{equation}
and
\begin{equation}
    \bar{A}
    =
    \frac{\sum_{i=1}^{N} w_i A_i}
         {\sum_{i=1}^{N} w_i}.
\end{equation}

\par
The uncertainty associated with the averaged value is evaluated from the
scatter of the measured values around the weighted average. Following
Ref.~\cite{Chen_2025}, this uncertainty is taken as the weighted standard
deviation,
\begin{equation}
    s(A)
    =
    \sqrt{
        \frac{
            \sum_{i=1}^{N} w_i (A_i-\bar{A})^2
        }{
            \frac{N-1}{N}\sum_{i=1}^{N} w_i
        }
    }.
\end{equation}

\par
When the individual values do not have assigned uncertainties, the
arithmetic mean is used,
\begin{equation}
    \bar{A}
    =
    \frac{1}{N}
    \sum_{i=1}^{N} A_i,
\end{equation}
and the uncertainty is evaluated using the standard deviation,
\begin{equation}
    s(A)
    =
    \sqrt{
        \frac{
            \sum_{i=1}^{N}(A_i-\bar{A})^2
        }{
            N-1
        }
    }.
\end{equation}
\par
In both cases, the averaged result is expressed as
\begin{equation}
    \langle A \rangle = \bar{A} \pm s(A).
\end{equation}

%
% Each of the commands below will create an unnumbered section with the appropriate heading.
% Remove any sections that are not relevant for your article.
% All sections except suppdata will be removed if the [anonymous] option is used.
% See iopjournal-guidelines.pdf for more information.
%

\ack{
%https://gwwiki.icrr.u-tokyo.ac.jp/JGWwiki/KAGRA/KSC/FAQ
This work was supported by MEXT, JSPS Leading-edge Research Infrastructure Program, JSPS Grant-in-Aid for Specially Promoted Research 26000005, JSPS Grant-in-Aid for Scientific Research on Innovative Areas 2402: 24103006, 24103005, and 2905: JP17H06358, JP17H06361 and JP17H06364, JSPS Core-to-Core Program A. Advanced Research Networks, JSPS Grant-in-Aid for Scientific Research (S) 17H06133, 20H05639 and 26K21721, JSPS Grant-in-Aid for Scientific Research (C) JP26K07123, JSPS Grant-in-Aid for Transformative Research Areas (A) 26A204: JP26H00407, the joint research program of the Institute for Cosmic Ray Research, University of Tokyo, National Research Foundation (NRF), Computing Infrastructure Project of Global Science experimental Data hub Center (GSDC) at KISTI, Korea Astronomy and Space Science Institute (KASI), and Ministry of Science and ICT (MSIT) in Korea, Academia Sinica (AS), AS Grid Center (ASGC) and the National Science and Technology Council (NSTC) in Taiwan under grants including the Science Vanguard Research Program, Advanced Technology Center (ATC) of NAOJ, Mechanical Engineering Center of KEK, and Vietnam National Foundation for Science and Technology Development (NAFOSTED) 103.01-2025.147.
}

%\funding{Sample text inserted for demonstration.}
% This section is a list of funder names and grant numbers

%\roles{Sample text inserted for demonstration.}
% List author names and the contributions made to the article, using terms from the NISO Contributor Roles Taxonomy (CRediT) https://credit.niso.org

%\data{Sample text inserted for demonstration.}
% For more information on IOP Publishing's research data policy see: https://publishingsupport.iopscience.iop.org/questions/research-data/

%\suppdata{Sample text inserted for demonstration.}

\bibliographystyle{mst-num}
\bibliography{cite}

\end{document}